# A diffusion MRI model for random walks confined on cylindrical surfaces: Towards non-invasive quantification of myelin sheath radius

Erick J Canales-Rodríguez[1,2,3,*], Chantal M.W. Tax[4,5], Elda Fischi-Gomez[2,1,3], Derek K. Jones[5], Jean-Philippe Thiran[3,1,2], Jonathan Rafael-Patiño[3,1]

[1] Department of Radiology, Centre Hospitalier Universitaire Vaudois (CHUV), Lausanne, Switzerland

[2] Computational Medical Imaging & Machine Learning Section, Center for Biomedical Imaging (CIBM), Lausanne, Switzerland

[3] Signal Processing Laboratory 5 (LTS5), Ecole Polytechnique Fédérale de Lausanne (EPFL), Lausanne, Switzerland.

[4] Image Sciences Institute, University Medical Center Utrecht, The Netherlands.

[5] Cardiff University Brain Research Imaging Centre (CUBRIC), Cardiff University, Cardiff, Wales, United Kingdom.

* Corresponding author: erick.canalesrodriguez@epfl.ch

**Abstract**

Quantifying the myelin sheath radius of myelinated axons in vivo is important for understanding, diagnosing, and monitoring various neurological disorders. Despite advancements in diffusion MRI (dMRI) microstructure techniques, there are currently no models specifically designed to estimate myelin sheath radii. This proof-of-concept theoretical study presents two novel dMRI models that characterize the signal from water diffusion confined to cylindrical surfaces, approximating myelin water diffusion. We derive their spherical mean signals, eliminating fiber orientation and dispersion effects for convenience. These models are further extended to account for multiple concentric cylinders, mimicking the layered structure of myelin. Additionally, we introduce a method to convert histological distributions of axonal inner radii from the literature into myelin sheath radius distributions. We also derive analytical expressions to estimate the effective myelin sheath radius expected from these distributions. Monte Carlo (MC) simulations conducted in cylindrical and spiral geometries validate the models. These simulations demonstrate agreement with analytical predictions. Furthermore, we observe significant correlations between the effective radii derived from histological distributions and those obtained by fitting the dMRI signal to a single-cylinder model. These models may be integrated with existing multi-compartment dMRI techniques, opening the door to non-invasive in vivo assessments of myelin sheath radii. Such assessments would require MRI scanners equipped with strong diffusion gradients, allowing measurements with short echo times. Further work is required to validate the technique with real dMRI data and histological measurements.

**Keywords**: Diffusion MRI; White matter microstructure, Myelin sheath radius; Myelin water; Monte Carlo simulations

## 1. Introduction

White matter (WM) primarily consists of axons [1], which are often enveloped by myelin produced by oligodendrocytes [2]. Myelin serves as an insulating sheath that enables nerve signals to propagate faster along the axon [3,4]. The axon-myelin unit interacts through complex molecular signaling and cellular processes, regulating the development and maintenance of myelin and the overall axon radius. Disruptions in the axon-myelin unit, such as demyelination or axon damage, are associated with neurological disorders such as multiple sclerosis [5], severe psychiatric conditions [6,7], and Alzheimer's disease [8]. These disorders are known to impair diverse cognitive functions [9]. Quantifying the microstructural properties of myelinated axons in vivo is crucial for enhancing our understanding of neurological diseases. This will ultimately improve diagnosis, early disease detection, and treatment of neurological disorders that affect millions worldwide.

Magnetic Resonance Imaging (MRI) is the primary technique for in vivo, non-invasive imaging of WM in the human brain. Many MRI techniques have been developed to characterize distinct WM properties [10–12]. For example, diffusion-weighted MRI (dMRI) measures the random motion of water molecules within and around axons. This sensitivity enables the estimation of spatial maps for various WM characteristics, such as axon orientations [13,14,23–30,15–22], dispersion [31,32], axon volume fraction [33–35], inner axon radii [10,11,43–46,12,36–42], intra- and extra-axonal water diffusivities [47,48], and T2 relaxation times [49,50]. In contrast, multi-echo T2 relaxometry [51,52,61,62,53–60] provides estimates closely correlated with myelin volume.

Despite considerable progress, challenges and research gaps remain in estimating the full range of WM microstructural features. One of them is the absence of specialized dMRI models explicitly designed for in vivo estimation of myelin sheath radii. Understanding water diffusion dynamics within myelin bilayers is essential, as the ‘apparent’ radial diffusivity of myelin water likely depends on the myelin sheath radius. This connection is promising, as it could enable myelin sheath radius estimation using dMRI data.

Accurately estimating myelin water diffusivities is challenging. This is because myelin water contributes minimally to the dMRI signal due to its short T2 time (i.e., 15 ms [52]), compared to the longer echo times (TE~80 ms) used in standard dMRI sequences. Nevertheless, various ex-vivo studies attempted to estimate myelin water diffusivities using T2 and T1 relaxation selective measurements. A diffusion-relaxation hybrid experiment proposed by [63], using a Carr-Purcell-Meiboom-Gill sequence, surprisingly revealed minor diffusional anisotropy and large parallel and radial diffusivities for the short T2 component associated with myelin water in the bovine optic nerve. Another approach employed T2 relaxation time to characterize myelin water selectively in the frog's peripheral nerve [64]. However, this ex-vivo study did not report myelin water diffusivities. On the other hand, T1 and T2 relaxation times have been utilized to observe myelin water in the excised frog sciatic nerve [65]. The T1-based method employed a double inversion recovery (DIR) sequence to nullify non-myelin water components, resulting in signals predominantly (>90%) derived from myelin water. This study found that myelin water diffusivities were lower when selected based on T1 characteristics with DIR-T1 measures (yielding parallel and radial diffusivities of $D_{\parallel}$=0.37-0.43 μm²/s and $D_{\perp}$=0.13-0.17 μm²/ms, respectively) compared to T2 characteristics ($D_{\parallel}$= 0.8 μm²/s and $D_{\perp}$= 0.19 μm²/ms).

Conversely, various in-vivo human brain studies have attempted to make the dMRI signal sensitive to the microstructure of myelin tissue. For instance, [66] implemented a magnetization transfer (MT) prepared stimulated-echo diffusion tensor imaging technique. The short TE=34 ms enabled by the stimulated-echo acquisition preserved a significant signal from the myelin water component with short T2, while the MT preparation further provided differentiating sensitization to this signal. Compared to the diffusion tensor derived from the conventional dMRI sequence acquired without MT preparation, the myelin water weighted tensor exhibited a significant increase in fractional anisotropy, most likely explained by the lower radial diffusivity of myelin water. In recent years, the diffusion-T2 relaxation approach has gained momentum thanks to the emergence of human scanners with strong diffusion gradients $G$ [67–69], allowing the use of diffusion sequences with shorter TEs. TE can be further reduced by using dMRI sequences with spiral readouts; for example, in the work by [70] and [71], TEs of 21.7 and 30 ms were achieved for $b$=1000 and 6000 s/mm² respectively, with $G$=300 mT/m, whereas [72] reduced the TE to 19 ms for $b$=1000 s/mm² with $G$=200 mT/m.

These recent studies suggest that it is possible to acquire dMRI data significantly weighted by myelin water. Therefore, this is an opportune time to develop new dMRI models for this often-overlooked WM compartment. In this theoretical and numerical proof of concept study, we propose a novel dMRI model for the diffusion process within a series of impermeable concentric cylinders separated by infinitesimal gaps filled with water, which could be employed as a first approximation to estimate myelin sheath radius. We derive the analytical dMRI signal and a Gaussian approximation with time-dependent radial diffusivity for this geometrical model and used Monte Carlo (MC) diffusion simulations to validate the proposed models.

This article is organized as follows. Section 2 presents our study's mathematical derivations, beginning with the geometrical model for the diffusion process in multiple concentric cylinders separated by infinitesimal distances (Subsection 2.1). We then model the dMRI signal as the product of signals generated by displacements parallel and perpendicular to the main cylinder's axis (Subsection 2.2) and introduce the diffusion propagator formalism to derive the analytical dMRI signal under the narrow-pulse approximation for pulsed gradient spin echo (PGSE) acquisitions (Subsection 2.3). A Gaussian approximation is presented in Subsection 2.4, followed by a refinement of these models in Subsection 2.5 to account for PGSE sequences with rectangular or trapezoidal diffusion gradients with non-narrow pulses. In Subsection 2.6, we derive the spherical mean signals, simplifying the modeling by eliminating fiber orientation and dispersion effects. In Subsection 2.7, we explore theoretical approximations to clarify how the estimated cylinder radius should be interpreted when fitting these models to measured data. The Methods section (Section 3) details the dMRI MC simulations designed to validate the proposed models. The results are presented in Section 4, followed by a discussion of their significance and the study's limitations in Section 5.

## 2. Theory

### 2.1 General description – geometrical model

Oligodendrocytes extend their cell membranes to wrap around axons in WM, creating multiple concentric layers of myelin. Each turn of wrapping adds another bilayer of myelin with a thickness of approximately $d_m$=4−5 nm. This process results in a multilayer spiral structure, with gaps of about $d_w$=3 nm thick [73] between the layers, filled by myelin water. Figure 1(A) shows a schematic transverse section of a myelinated axon.

In this study, we approximate the diffusion process along this spiral trajectory as diffusion within a series of impermeable concentric solid cylinders separated by infinitesimal water-filled gaps (see Figure 1(B)). The rationale for this approximation is as follows: For a given diffusion time, a diffusing water molecule traveling a total displacement of $2\pi aN$ (where $a$ is the myelin radius at the starting position and $N$ is an arbitrary number) along the spiral trajectory experiences a net radial displacement of about $N(d_w+d_m)$ (see cross-sectional plane shown in Figure 1(A). This displacement remains negligible, even for molecules traveling long distances. For example, for $a$=0.5 µm and $N$=10, the path length along the spiral is 31.4 µm, and the net radial displacement is approximately 0.08 µm, hence significantly smaller than the minimum displacement required to attenuate the dMRI signal in state-of-the-art scanners [39,74]. Moreover, since spin echo dMRI sequences designed to be sensitive to myelin water employ short TEs (equivalently short diffusion times), most molecules will travel relatively short distances along the spiral trajectory, minimizing the net radial displacement.

For this reason, we propose to simplify the spiral trajectory by using concentric cylinders of similar size. As infinitesimal distances separate the cylinders, we assume that the underlying diffusion process is equivalent to random walks confined to the cylinder surfaces. Therefore, we will first derive the diffusion propagator for Brownian motion on the cylinder surface, see Figure 1(C), and then extend this model to multiple cylinders. Moreover, to eliminate fiber orientation and dispersion effects (confounding factors), we will derive the spherical mean dMRI signal for this model. This approach will help us to interpret the mean radius estimated by fitting a single-cylinder-surface model to the dMRI signal arising from multiple cylindrical surfaces.

Insert Figure 1 around here (1.5 columns)

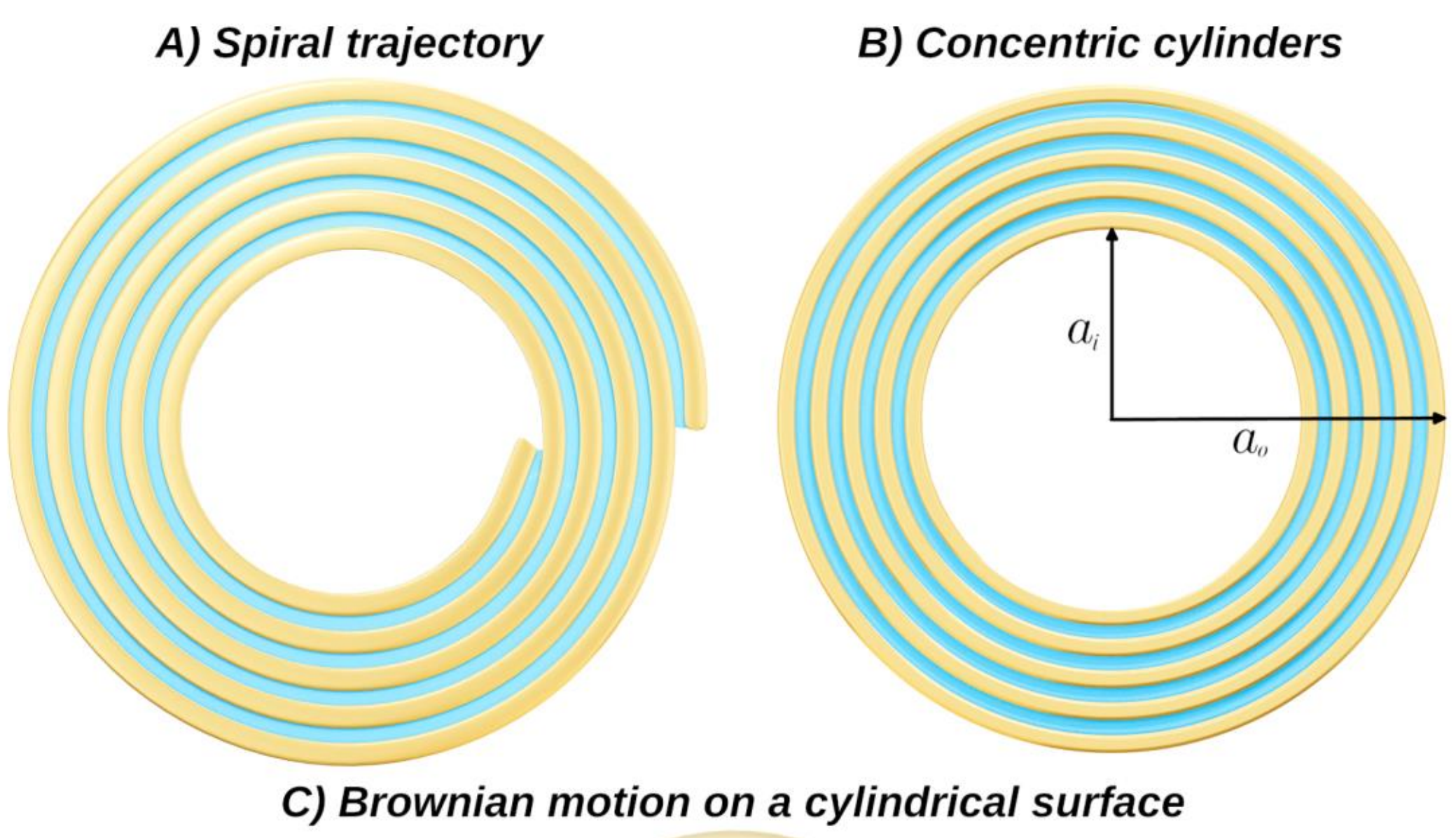

**Figure 1. Schematic representation of an axon and its myelin sheath.** (A) Cross-sectional view of a myelinated axon showing the spiral trajectory of compact myelin bilayers (in yellow-orange). Each myelin bilayer has a thickness of approximately 4-5 nm and is separated by myelin water gaps (i.e., cytoplasmic and extracellular water) (in blue) with a thickness of approximately 3 nm [73]. (B) Cross-section of multiple concentric alternating cylinders representing the myelin bilayers and myelin water. This simplified geometrical model is used to study the diffusion process. (C) Example of myelin water molecules (represented by blue dots) diffusing on a cylindrical surface, where $a$ represents the radius and $\Phi$ denotes the polar angle, quantifying the displacement of a water molecule along the 2D surface in the x-y plane. This plane is assumed to be perpendicular to the main axis of the cylinder, which is oriented along the z-axis.

### 2.2 Decoupling diffusive motions

To simplify our model, we will consider an infinitely long cylinder whose main axis is oriented along the z-axis, with its transverse section lying in the x-y plane. An important aspect of this model is that the dMRI signal can be decomposed into contributions from spin particles diffusing parallel and perpendicular to the cylinder's main axis. In this coordinate frame of reference, these diffusion processes are statistically independent. Therefore, the displacement probability distribution $P(\mathbf{r},t) = P(\mathbf{r}_{xy},t)P(\mathbf{r}_z,t)$ can be expressed as the product of the distributions for motion in the perpendicular $P(\mathbf{r}_{xy},t)$ and parallel $P(\mathbf{r}_z,t)$ directions. The net displacement vector $\mathbf{r} = \mathbf{r}_{xy} + \mathbf{r}_z$ at diffusion time $t$ can be decomposed into the displacement vectors perpendicularly and parallel to the cylinder's axis. Note that $\mathbf{r}_{xy} = r_x\hat{\mathbf{i}} + r_y\hat{\mathbf{j}}$ and $\mathbf{r}_z = r_z\hat{\mathbf{k}}$, where $r_x$, $r_y$ and $r_z$ are the vector's lengths along the unit vectors $\hat{\mathbf{i}}$, $\hat{\mathbf{j}}$, $\hat{\mathbf{k}}$ associated with the x-, y-, and z-axes, respectively.

For this type of decoupled diffusive motion, [75] showed that the dMRI signal can be expressed as the product of the dMRI signals arising from displacement parallel and perpendicular to the cylinder's axis: $E(\mathbf{q},t) = E_{\perp}(\mathbf{q}_{xy},t)E_{\parallel}(\mathbf{q}_z,t)$, where $\mathbf{q}_{xy} = q_x\hat{\mathbf{i}} + q_y\hat{\mathbf{j}}$ and $\mathbf{q}_z = q_z\hat{\mathbf{k}}$, $\mathbf{q} = \mathbf{q}_{xy} + \mathbf{q}_z = \gamma\mathbf{g}\delta$, $\gamma$ is the gyromagnetic ratio of the diffusing spin particles (e.g., hydrogen nuclei), $\mathbf{g} = G\hat{\mathbf{g}}$ denotes the applied diffusion gradient with magnitude $G$ and unit orientation vector $\hat{\mathbf{g}}$, and $\delta$ is the duration of the diffusion gradient pulses. Note that $t$ should be expressed in terms of the dMRI sequence time parameters. A general detailed derivation of this decoupled signal model is provided in [75].

### 2.3 Diffusion MRI signal and diffusion propagator: narrow-delta approximation

In this section, we will derive the analytical expressions for $E_{\|}\left(\mathbf{q}_z,t\right)$ and $E_{\perp}\left(\mathbf{q}_{xy},t\right)$ necessary to provide the full dMRI signal model. This derivation follows the diffusion propagator formalism under the narrow-pulse (narrow-delta) approximation, which assumes that the duration of the diffusion gradient is very short ($\delta \to 0$). Thus, under this formalism and for a pulsed-gradient spin echo (PGSE) sequence [76], the diffusion time is equal to the time difference between the onset of the two diffusion gradients $t=\Delta$.

The dMRI signal $E_{\|}\left(\mathbf{q}_z,t\right)$ arising from displacements parallel to the cylinder's main axis $\mathbf{r}_z$ is related to the 1D displacement probability distribution by the following Fourier-relationship:

$$\frac{E_{\|}\left(\mathbf{q}_z,t\right)}{E_{\|}\left(\mathbf{q}_z=0,t\right)}=\int_{-\infty}^{\infty}\int_{-\infty}^{\infty} P\left(\mathbf{z}\middle|,\mathbf{z}',t\right)P\left(\mathbf{z}'\right)e^{iq_z(z-z')}dz'dz\,, \tag{1}$$

where $P\left(\mathbf{z}'\right)$ is the probability for a particle to be at position $\mathbf{z}'=z'\hat{\mathbf{k}}$ at initial time $t=0$, and $P\left(\mathbf{z}\middle|,\mathbf{z}',t\right)$ is the probability that a particle initially located at position $\mathbf{z}'$ migrate to position $\mathbf{z}=z\hat{\mathbf{k}}$ in time $t$. Assuming that at $t=0$ all particles are uniformly distributed along the cylinder's axis (i.e., $P\left(\mathbf{z}'\right)$ is constant) and using the change of variables $\mathbf{r}_z=\mathbf{z}-\mathbf{z}'$ to quantify displacements, Eq. (1) can be rewritten as

$$\frac{E_{\|}\left(\mathbf{q}_z,t\right)}{E_{\|}\left(\mathbf{q}_z=0,t\right)}=\int_{-\infty}^{\infty} P\left(\mathbf{r}_z,t\right)e^{iq_z r_z}dr_z\,, \tag{2}$$

Since the motion of particles along the cylinder's main axis is unrestricted (assuming an infinitely long cylinder), we assume 1D Gaussian diffusion with a characteristic myelin water diffusivity $D_{\|}$ on the cylinder's surface:

$$P\left(\mathbf{r}_z,t\right)=\frac{1}{\sqrt{4\pi D_{\|}t}}e^{-\frac{r_z^2}{4D_{\|}t}}. \tag{3}$$

The resulting integral is solved, obtaining the familiar dMRI signal expression for Gaussian diffusion,

$$E_{\|}\left(\mathbf{q}_z,t\right)=E_{\|}\left(\mathbf{q}_z=0,t\right)e^{-q_z^2D_{\|}t}. \tag{4}$$

Likewise, the dMRI signal arising from displacements perpendicular to the cylinder's axis $E_{\perp}\left(\mathbf{q}_{xy},t\right)$ depends on the 2D displacement probability distribution by the following Fourier-relationship:

$$\frac{E_{\perp}\left(\mathbf{q}_{xy},t\right)}{E_{\perp}\left(\mathbf{q}_{xy}=0,t\right)}=\int\limits_{\mathbb{R}^2}\int\limits_{\mathbb{R}^2}P\left(\mathbf{r}_{xy}\middle|,\mathbf{r}'_{xy},t\right)P\left(\mathbf{r}'_{xy}\right)e^{i\mathbf{q}_{xy}\left(\mathbf{r}_{xy}-\mathbf{r}'_{xy}\right)}d\mathbf{r}'_{xy}d\mathbf{r}_{xy}\,, \tag{5}$$

where $P\left(\mathbf{r}'_{xy}\right)$ and $P\left(\mathbf{r}_{xy}\middle|,\mathbf{r}'_{xy},t\right)$ are the probability of finding a particle at position $\mathbf{r}'_{xy}$ in the x-y plane at $t=0$, and the probability of moving from $\mathbf{r}'_{xy}$ to $\mathbf{r}_{xy}$ in time $t$.

As the particle displacements in the plane perpendicular to the cylinder's axis are confined on a circle, it is convenient to rewrite the integrals in Eq. (5) in polar coordinates due to the polar symmetry of this system,

$$\begin{aligned}\frac{E_{\perp}\left(\mathbf{q}_{xy},t\right)}{E_{\perp}\left(\mathbf{q}_{xy}=0,t\right)}=&\int\limits_0^{\infty}\int\limits_0^{\infty}\int\limits_0^{2\pi}\int\limits_0^{2\pi}P\left(\rho,\theta\middle|,\rho',\theta',t\right)P\left(\rho',\theta'\right)\times\\&e^{iq_{xy}\rho\cos\left(\varphi+\theta\right)}e^{-iq_{xy}\rho'\cos\left(\varphi+\theta'\right)}\rho\rho'd\rho'd\rho d\theta d\theta',\end{aligned} \tag{6}$$

where the cartesian components of the 2D vectors, $\mathbf{r}_{xy}$, $\mathbf{r}'_{xy}$, $\mathbf{q}_{xy}$, are rewritten in terms of their magnitudes, $\rho$, $\rho'$, $q_{xy}$, and angles of orientation, $\theta,\theta',\varphi$, respectively: $\mathbf{r}_{xy}=\left(\rho\cos\left(\theta\right),\rho\sin\left(\theta\right)\right)$, $\mathbf{r}'_{xy}=\left(\rho'\cos\left(\theta'\right),\rho'\sin\left(\theta'\right)\right)$, and $\mathbf{q}_{xy}=\left(q_{xy}\cos\left(\varphi\right),q_{xy}\sin\left(\varphi\right)\right)$.

In **Appendix A**, we show that Eq. (6) can be simplified to

$$\frac{E_{\perp}\left(\mathbf{q}_{xy},t\right)}{E_{\perp}\left(\mathbf{q}_{xy}=0,t\right)}=\frac{1}{2\pi}\int\limits_0^{2\pi}\int\limits_0^{2\pi}P\left(\Phi\middle|,a,t\right)e^{iq_{xy}a\cos\left(\psi\right)}e^{-iq_{xy}a\cos\left(\psi-\Phi\right)}d\psi d\Phi, \tag{7}$$

where we used the change of variables $\psi=\varphi+\theta$ and $\Phi=\theta-\theta'$, and $P\left(\Phi\middle|,a,t\right)$ is the probability that the particles' motion on the circle with radius $a$ covers a polar angle $\Phi$ at a time $t$, see Figure 1 (C).

We model $P\left(\Phi\middle|,a,t\right)$ as a wrapped Gaussian distribution [77] with diffusivity $D$:

$$P\left(\Phi\middle|,a,t\right)=\frac{1}{\sqrt{4\pi Dt}}\sum_{p=-\infty}^{\infty}e^{-\frac{a^2}{4Dt}(\Phi+2\pi p)^2}$$
$$=\frac{1}{2\pi}\left[1+2\sum_{p=1}^{\infty}e^{-p^2\frac{Dt}{a^2}}\cos\left(p\Phi\right)\right]. \tag{8}$$

This distribution results from wrapping the 1D Gaussian distribution (on the infinite line) around the circle's circumference. It takes into account that during the diffusion process, a population of particles could travel distances larger than $2\pi ap$, where $2\pi a$ is the perimeter of the circle, and $p=1,2,\ldots,\infty$. The second expression in Eq. (8) provides a helpful alternative representation of this function [77–79]. It is the solution of the diffusion equation of Brownian particles confined in a circle $S^1$ [80–82]. However, note that in [81], the function was normalized with the circle's perimeter, whereas our distribution is normalized with the angle, i.e., $\int_0^{2\pi}P\left(\Phi\middle|,a,t\right)d\Phi=1$.

Assuming that the translational diffusion parallel to the cylinder's main axis and along the "unwrapped" circle are equal, then $D=D_{\|}$. After substituting Eq. (8) into Eq. (7) we obtain

$$\frac{E_{\perp}\left(\mathbf{q}_{xy},t\right)}{E_{\perp}\left(\mathbf{q}_{xy}=0,t\right)}=J_0^2\left(aq_{xy}\right)+2\sum_{p=1}^{\infty}J_p^2\left(aq_{xy}\right)e^{-p^2\frac{D_{\|}t}{a^2}}, \tag{9}$$

where $J_p$ is the *p-th* Bessel function of the first kind. The complete derivation is shown in **Appendix A**. This expression does not depend on the orientation $\varphi$ of vector $\mathbf{q}_{xy}$ in the plane perpendicular to the cylinder's axis due to transverse symmetry, as expected. In the limit $D_{\|}t>>a^2$, Eq. (8) becomes a uniform distribution and Eq. (9) tends to

$$\frac{E_{\perp}\left(\mathbf{q}_{xy},t>>a^2/D\right)}{E_{\perp}\left(\mathbf{q}_{xy}=0,t>>a^2/D\right)}\approx J_0^2\left(aq_{xy}\right), \tag{10}$$

which does not depend on $t$.

An independent derivation of Eq. (9) was reported in [64,83]. However, the result reported by [64] was obtained by assuming a Gaussian distribution for the angular motion instead of a Wrapped Gaussian, which solution only tends to Eq. (9) in the limit case when $a^2>>D_{\|}t$.

By merging results from Eqs. (4) and (9), we obtain the final signal model for a single cylinder:

$$E(\mathbf{q},t) = E_{\parallel}(\mathbf{q}_z,t)\,E_{\perp}(\mathbf{q}_{xy},t)$$
$$= E(\mathbf{q}=0,t)\,e^{-q^2\cos(\beta)^2 D_{\parallel}t}\left[J_0^2\left(aq\sin(\beta)\right) + 2\sum_{p=1}^{\infty} J_p^2\left(aq\sin(\beta)\right)e^{-p^2\frac{D_{\parallel}t}{a^2}}\right], \qquad (11)$$

where $\beta$ is the angle between the diffusion gradient orientation and the cylinder's axis, $q_{xy} = q\sin(\beta)$ and $q_z = q\cos(\beta)$. For practical purposes, the signal can be adequately approximated by the first $p = 1,\ldots,P$ terms in the series.

## 2.4 Gaussian approximation

When the displacement probability distribution in the x-y plane (perpendicular to the cylinder's axis) is approximated by an isotropic bivariate Gaussian distribution, the mean-squared displacement of particles $\left\langle \left|\mathbf{r}_{xy}\right|^2 \right\rangle$ is related to the 'apparent' radial diffusivity in the 2D plane according to $D_{\perp}^{app} = \left\langle \left|\mathbf{r}_{xy}\right|^2 \right\rangle / 4t$. For such an isotropic Gaussian diffusion process, the corresponding dMRI signal $E_{\perp}(\mathbf{q}_{xy},t)$ is given by

$$E_{\perp}(\mathbf{q}_{xy},t) = E_{\perp}(\mathbf{q}_{xy}=0,t)\,e^{-q_{xy}^2 D_{\perp}^{app} t}. \qquad (12)$$

The expression for $D_{\perp}^{app}$ depends on the diffusion time and circle radius $a$ as

$$D_{\perp}^{app} = \frac{a^2}{2t}\left[1 - \mathrm{e}^{-\frac{D_{\parallel}t}{a^2}}\right], \qquad (13)$$

where we assumed $D = D_{\parallel}$, like in Eq. (9). The full derivation is presented in **Appendix B**. For very short diffusion times, $t \to 0$, the apparent radial diffusivity does not depend on the circle's radius, $D_{\perp}^{app} = D_{\parallel}/2$, because no structural features are probed at such a small time-scale. Conversely, for very long diffusion times, $D_{\perp}^{app} \to a^2/2t$.

The final dMRI signal, considering both the parallel and radial diffusion components, is given by

$$E(\mathbf{q},t) = E(\mathbf{q}=0,t)\,e^{-q^2\cos(\beta)^2 D_{\parallel}t}\,e^{-q^2\sin(\beta)^2 D_{\perp}^{app}t}. \qquad (14)$$

This analytical form is equivalent to an axially symmetric diffusion tensor signal, as described in Eq. (5) in [84]. However, note that the radial diffusivity depends on the diffusion time and the size of the confining geometry, i.e., the cylinder radius.

### 2.5 Correction for non-narrow deltas

Our previous derivations are based on the q-space formalism (see Eqs. (1) and (5)). This approach is valid for PGSE sequences [76] using diffusion-encoding gradients with infinitesimal duration $\delta$. Consequently, the proposed signal models are not valid for sequences that do not fulfill this requirement. In this section, we will use the q-space correction approach presented by [85] to provide more general signal approximations beyond this acquisition protocol.

Under the narrow pulse approximation, the dephasing of the spins due to their motion during the application of the diffusion gradients is neglected. Thus, the diffusion time is equal to the time difference between the onset of the two diffusion gradients. However, for finite $\delta$ it is unclear what diffusion time derived from the PGSE sequence must be used in the diffusion propagator to evaluate the dMRI model. This problem was tackled by [85], who proposed a general relationship between the signal attenuation $\left\langle e^{i\phi} \right\rangle_{\Delta,\delta,\mathbf{g}}$ for the PGSE sequence and the displacement probability

$$\frac{\left\langle e^{i\phi} \right\rangle_{\Delta,\delta,\mathbf{g}}}{\left\langle e^{i\phi} \right\rangle_{\Delta,\delta,\mathbf{g}=0}} = \int_{\mathbb{R}^3} P\left(\mathbf{r}, t_{\exp}\right) \left\langle e^{i\phi} \middle| \mathbf{r} \right\rangle_{\Delta,\delta,\mathbf{g}} d\mathbf{r}, \tag{15}$$

where the integral is over the infinite three-dimensional space, $t_{\exp}$ is the total diffusion time of the experiment between the onset of the first gradient and the termination of the second gradient, and $\left\langle e^{i\phi} \middle| \mathbf{r} \right\rangle_{\Delta,\delta,\mathbf{g}}$ denotes the average signal attenuation (dephasing) of the population of spins experiencing a net displacement $\mathbf{r}$ in time $t_{\exp}$. Note that $t_{\exp} = \Delta + \delta$ for PGSE sequences with rectangular diffusion gradients and $t_{\exp} = \Delta + \delta + \xi$ for trapezoidal diffusion gradients, where $\xi$ is the rise time of the trapezoidal ramp [86].

In **Appendix C**, we provide a compact re-derivation of Lori's approach, which found the following approximation:

$$\frac{\left\langle e^{i\phi} \right\rangle_{\Delta,\delta,\mathbf{g}}}{\left\langle e^{i\phi} \right\rangle_{\Delta,\delta,\mathbf{g}=0}} \approx \int_{\mathbb{R}^3} P\left(\mathbf{r}, t_{\exp}\right) e^{i\mathbf{q}'\mathbf{r}} d\mathbf{r}, \tag{16}$$

where $\mathbf{q}' = \mathbf{q}\sqrt{t_{eff}/t_{\exp}}$ is a scaled q-space vector; $t_{eff}$ denotes the 'effective' diffusion time that appears in the $b$-value definition, i.e., $b = q^2 t_{eff}$, which is equal to $t_{eff} = \Delta - \delta/3$ and $t_{eff} = \Delta - \delta/3 + \xi^3/30\delta^2 - \xi^2/6\delta$ for rectangular and trapezoidal diffusion gradients, respectively [86]. According to this result, the q-space formalism can still be employed to relate the diffusion propagator and the dMRI signal attenuation produced by a PGSE sequence with finite $\delta$. However, it must be corrected by evaluating the diffusion propagator at the total diffusion encoding time $t_{\exp}$ and using a modified q-space vector $\mathbf{q}'$. Note that for narrow pulses, the correction converges to the classical q-space formalism with $t_{\exp} = t_{eff} = \Delta$, and $\left\langle e^{i\phi} \middle| \mathbf{r} \right\rangle_{\Delta,\delta,\mathbf{g}} = e^{i\mathbf{qr}}$, as expected.

The theoretical result in Eq. (16) was confirmed in [85] by numerical simulations for homogeneous Gaussian diffusion, heterogeneous diffusion in permeable microscopic Gaussian domains, and diffusion inside restricted spherical reflecting domains. In all the analyses, this correction produced better results than using the original q-vector and the relationship $t_{\exp} = \Delta - \delta/3$, for rectangular diffusion gradients. It is important to notice that this approach may only provide a precise correction for displacement distributions that do not deviate significantly from a Gaussian distribution.

In this study, we will use this correction to evaluate our signal models in Eqs. (11) and (14).

### 2.6 Spherical mean signals

The previous signal models, see Eqs. (11) and (14), are based on the assumption of a single cylindrical surface. In the case of a distribution of cylinders with equal radius but multiple orientations, the orientation effect can be removed from Eq. (11) by computing the orientation-averaged spherical mean signal $\langle E \rangle$. Following the approach of [33,87–89], we obtain,

$$
\begin{aligned}
\frac{\left\langle E\left(q,\Delta,\delta,\xi,a\right)\right\rangle}{E\left(q=0\right)} = \frac{1}{2}\Bigg[ & \sum_{k=0}^{\infty}\sum_{j=0}^{k}\frac{\left(-1\right)^{k}}{\left(k!\right)^{2}}\binom{2k}{k}\left(\frac{aq}{2}\sqrt{\frac{t_{eff}}{t_{\exp}}}\right)^{2k} \\
& \times\binom{k}{j}\left(-1\right)^{j}\frac{\Gamma\left(j+\frac{1}{2}\right)-\Gamma\left(j+\frac{1}{2},q^{2}D_{\|}t_{eff}\right)}{\left(q^{2}D_{\|}t_{eff}\right)^{j+1/2}} \\
& +\sum_{p=1}^{\infty}\sum_{k=0}^{\infty}\sum_{j=0}^{p+k}e^{-p^{2}\frac{D_{\|}t_{\exp}}{a^{2}}}\frac{\left(-1\right)^{k}}{k!\left(2p+k\right)!}\binom{2\left(p+k\right)}{p+k}\left(\frac{aq}{2}\sqrt{\frac{t_{eff}}{t_{\exp}}}\right)^{2\left(p+k\right)} \\
& \times\binom{p+k}{j}\left(-1\right)^{j}\frac{\Gamma\left(j+\frac{1}{2}\right)-\Gamma\left(j+\frac{1}{2},q^{2}D_{\|}t_{eff}\right)}{\left(q^{2}D_{\|}t_{eff}\right)^{j+1/2}}\Bigg],
\end{aligned}
\tag{17}
$$

where $t_{eff}$ and $t_{exp}$ depend on the experimental parameters $\{\Delta,\delta,\xi\}$.

A detailed derivation of this expression is presented in **Appendix D**, which also includes Lori's q-space correction described in the previous section.

On the other hand, for the Gaussian diffusion model in Eq. (14) with time-dependent radial diffusivity, the spherical mean signal is equivalent to that from an axis-symmetric diffusion tensor [33,37,43,84,87,90]:

$$
\frac{\left\langle E\left(q,\Delta,\delta,\xi,a\right)\right\rangle}{E\left(q=0\right)} = \sqrt{\frac{\pi}{4}}e^{-bD_{\perp}^{app}}\frac{erf\left(\sqrt{b\left(D_{\|}-D_{\perp}^{app}\right)}\right)}{\sqrt{b\left(D_{\|}-D_{\perp}^{app}\right)}}\ , \tag{18}
$$

where $erf$ denotes the error function. In our model, the radial diffusivity $D_{\perp}^{app}(a,t=\Delta+\delta)$ depends on the cylinder radius $a$ and the total diffusion time according to the model defined in Eq. (13) and incorporating Lori's correction. Note that this correction does not affect the *b*-value since $b=q^{2}t_{eff}=q'^{2}t_{\exp}$ for rectangular and trapezoidal diffusion gradients.

### 2.7 Estimating the mean myelin sheath radius: what do we measure?

In this section, we will derive the spherical mean dMRI signal for a distribution of cylinders with different radii. Specifically, we will consider two cases:

1. *Multiple concentric cylinders*: This model represents the diffusion process of myelin water within a single axon. Each cylinder corresponds to a layer of the myelin sheath. The diffusion process is confined to these cylindrical surfaces, and the overall dMRI signal is the sum of contributions from each cylindrical layer; see Figure 1 (B).
2. *Distribution of multiple concentric cylinders with different radii*: This model represents a voxel with multiple axons, where the inner axon radius follows a Gamma distribution. The Gamma distribution is a flexible choice that can model a wide range of axon radius distributions observed in neural tissues [40]; see Figure 2.

We aim to define and estimate the 'effective' myelin sheath radius by approximating the signal from multiple cylindrical surfaces with the signal from a single cylindrical surface. The effective myelin sheath radius simplifies the complex distribution into a single representative value. This approach is analogous to axon diameter mapping techniques, which estimate an effective radius from an underlying distribution of inner axon radii [10,11,36–41,45].

### 2.7.1 Effective myelin sheath radius for a single axon

The spherical mean dMRI signal $\langle E_{axon} \rangle$ arising from $N$ concentric cylindrical surfaces is given by

$$
\begin{aligned}
\frac{\langle E_{axon}(q,\Delta,\delta,\xi) \rangle}{E_{axon}(q=0)} &= \frac{1}{E_{axon}(q=0)} \sum_{i=1}^{N} \langle E(q,\Delta,\delta,\xi,a_i) \rangle \\
& \sum_{i=1}^{N} \frac{E(q=0,a_i)}{E_{axon}(q=0)} \langle S(q,\Delta,\delta,\xi,a_i) \rangle, \\
&= \sum_{i=1}^{N} \frac{a_i}{\left( \sum_{j=1}^{N} a_j \right)} \langle S(q,\Delta,\delta,\xi,a_i) \rangle.
\end{aligned}
\tag{19}
$$

where the summation is over all cylinder's radii from the inner radius $a_1$ to the outer radius $a_N$, and $\langle S(q) \rangle = \langle E(q) \rangle / E(q=0)$ denotes the spherical mean dMRI signal produced by each cylinder normalized by its baseline signal without diffusion weighting ($q$=0 image); see Eqs. (17) and (18). The term $E_{axon}(q=0)$ was included on both sides of the equation on purpose. Since $E(q=0)$ is proportional to the number of diffusing spin particles, $E(q=0,a_i)/E_{axon}(q=0)$ is the ratio of the number of those particles on the cylinder with the radius $a_i$ and the total number on all cylinders. Assuming the same proton density (i.e., number of particles per unit surface area) and cylinder length, this ratio is the surface area of the *i*-th cylinder divided by the total surface area of all cylinders, or equivalently, the radius of the *i*-th cylinder divided by the sum of all radii.

We can substitute the normalized spherical mean signal obtained for the general model (Eq. (17)) or the Gaussian approximation with time-dependent radial diffusivity (Eq. (18)) in Eq. (19). When the resulting signal is approximated by the signal from a single cylindrical surface, then

$$\frac{\left\langle E\left(q,\Delta,\delta,\xi,a_{eff}\right)\right\rangle}{E\left(q=0\right)} \approx \sum_{i=1}^{N} \frac{a_i}{\left(\sum_{j=1}^{N} a_j\right)} \left\langle S\left(q,\Delta,\delta,\xi,a_i\right)\right\rangle, \tag{20}$$

where $a_{eff}$ is the effective radius. This effective radius represents the MRI-visible radius that considers that the measured signal is weighted by the radius, such that the outer cylinder contributes more than the inner cylinder to the measured data. Assuming that all cylinders have the same distance between them, then $a_{eff}$ will be more biased towards $a_N$ than towards $a_1$ from the arithmetic mean $a_{eff} \geq \langle a \rangle$, defined by

$$\begin{aligned} \langle a \rangle &= \frac{1}{N}\sum_{i=1}^{N} a_i, \\ &= \frac{1}{N}\sum_{i=1}^{N}\left(a_1 + \left(i-1\right)\Delta a\right), \\ &= a_1 + \left(\frac{N-1}{2}\right)\Delta a, \\ &= \frac{a_i + a_o}{2}, \end{aligned} \tag{21}$$

where $\Delta a = a_{i+1} - a_i$ is the distance between two consecutive cylinders, and the outer cylinder's radius is $a_N = a_1 + \left(N-1\right)\Delta a$. In the previous equation, we replaced $a_1$ and $a_N$ with the inner and outer axon radii, $a_i$, $a_o$, respectively.

### 2.7.2 Effective myelin sheath radius for a distribution of axon radii

For a sample of myelinated axons with the same g-ratio, $g = a_i / a_o$, and the distribution of inner axon radius parameterized by $P\left(a_i\right)$, the marginal distribution of myelin sheath (cylinder) radii is given by

$$P(a) = \eta \int_0^\infty U\left(a \middle| a_i, a_o\right) P\left(a_i\right) da_i , \tag{22}$$

where $\eta$ is the normalization constant ensuring that $\int_0^\infty P\left(a\right) da = 1$, and $U\left(a \middle| a_i, a_o\right)$ is a uniform distribution modeling the myelin layers of each axon as uniformly distributed cylinders in the interval $\left[a_i, a_o\right]$,

$$\begin{aligned} U\left(a \middle| a_i, a_o\right) &= \frac{1}{a_o - a_i} \mathbf{1}_{[a_i, a_o]}(a), \\ &= \frac{g}{a_i\left(1-g\right)} \mathbf{1}_{[a_i, a_i/g]}(a), \end{aligned} \tag{23}$$

which is written in terms of $a_i$ and $g$. The indicator function $\mathbf{1}_{[a_i, a_o]}(a)$ is equal to 1 if $a_i \le a \le a_o$ and 0 otherwise.

We assume a Gamma distribution for the inner radius as in [40]:

$$P\left(a_i\right) = \frac{\kappa^\mu}{\Gamma\left(\mu\right)} a_i^{\mu-1} e^{-\kappa a_i} , \text{ for } a_i > 0,\ \mu, \kappa > 0, \tag{24}$$

where $\Gamma\left(\mu\right)$ is the Gamma function, and $\mu$ and $\kappa$ are the shape and inverse scale parameters, respectively, such that the mean radius and variance are $\left\langle a \right\rangle = \mu/\kappa$ and $\sigma^2 = \mu/\kappa^2$ .

Inserting Eqs. (23) and (24) into Eq. (22), and considering that at a given radius $a$ only those cylinders in the range from $\left[a \cdot g, a/g\right]$ contribute to the integral (i.e., the population of cylinders from axons with inner and outer radii ranging from [ $a_i = a \cdot g$ , $a_o = a$ ] to [ $a_i = a$ , $a_o = a/g$ ]), we obtain

$$\begin{aligned} P(a) &= \eta \frac{\kappa^\mu}{\Gamma\left(\mu\right)} \frac{g}{\left(1-g\right)} \int_{a \cdot g}^{a/g} a_i^{\mu-2} e^{-\kappa a_i} da_i , \\ &= \frac{\kappa}{\Gamma\left(\mu\right)} \frac{g}{\left(1-g^2\right)} \left[ \Gamma\left(\mu-1, a \cdot g \cdot \kappa\right) - \Gamma\left(\mu-1, \frac{a \cdot \kappa}{g}\right) \right], \end{aligned} \tag{25}$$

where $\Gamma\left(s,x\right)$ denotes the upper incomplete Gamma function. The complete derivation is developed in **Appendix E**. Note that for axons with a very small number of myelin layers, $g \to 1$ and $P(a) \approx P\left(a_i\right)$. Figure 2 shows an example of a distribution of inner axon radius sampled from the splenium of the corpus callosum of a human brain reported by [45,91] and the corresponding marginal distribution of myelin sheath radii assuming $g = 0.6$.

The spherical mean dMRI signal produced by such a distribution of cylinders is

$$\frac{\langle E_{dist}(q,\Delta,\delta,\xi)\rangle}{E_{dist}(q=0)} = \frac{\int_0^\infty aP(a)\langle S(q,\Delta,\delta,\xi,a)\rangle da}{\int_0^\infty aP(a)da}, \tag{26}$$

If the distribution of the inner radius $P(a_i)$ and the g-ratio are known from histological measurements, we can estimate $P(a)$ from Eq. (25). The dMRI signal in Eq. (26) can be computed numerically using Eq. (17) or Eq. (18) for a given set of PGSE acquisition parameters, and the effective radius $a_{eff}$ can then be estimated by fitting the single-cylinder model to the resulting signal.

Insert Figure 2 around here (2 columns)

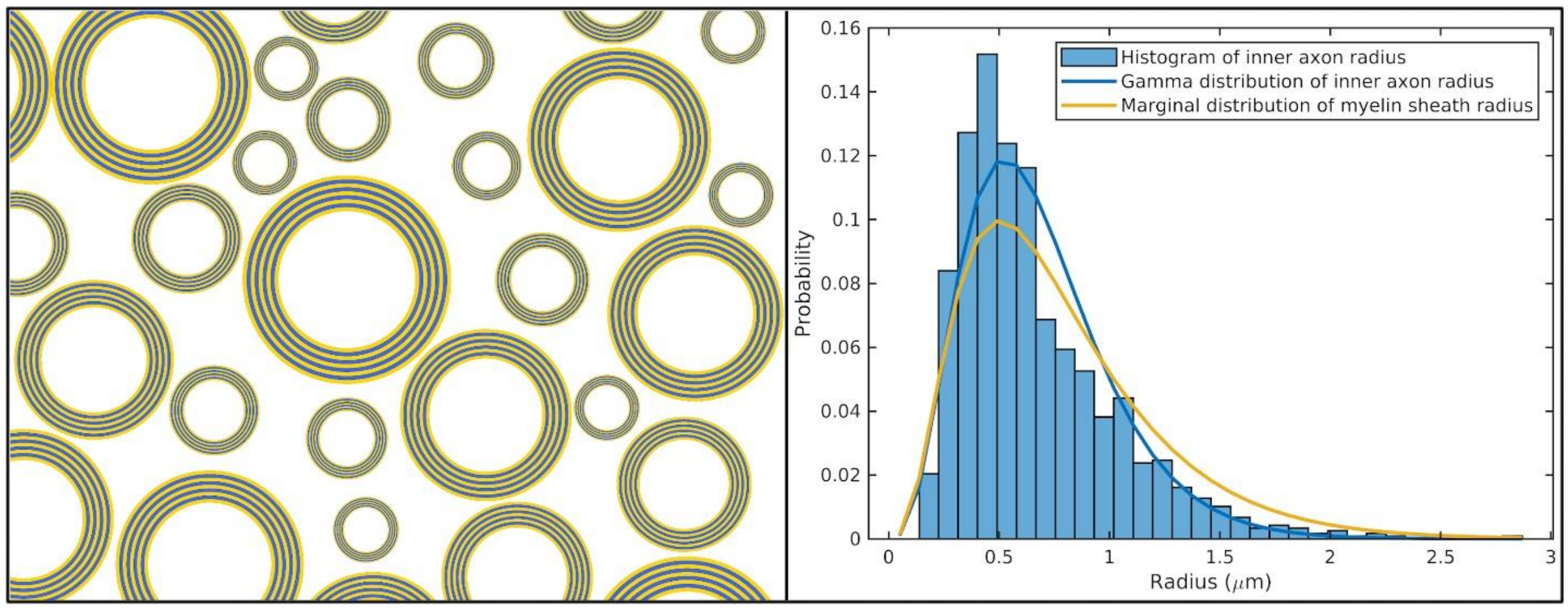

**Figure 2. Distribution of radius.** Left Panel: The diagram illustrates a population of axons within a voxel, displaying varying inner radii while maintaining a constant g-ratio. Right Panel: This graph presents the distribution of inner axon radii sampled from the splenium of the Corpus Callosum of an ex-vivo human brain (data from [91]). The Gamma distribution fitting the measured inner radii is depicted in blue, and the corresponding marginal distribution of the myelin sheath radius calculated using Eq. (25) and assuming a constant g-ratio of 0.6, is shown in yellow-orange. The Gamma distribution was fitted to the data using a Maximum Likelihood approach, as implemented in the *gamfit* function in @Matlab. This visualization highlights the relationship between the inner axon radius distribution (mean=0.68 µm, variance=0.11 µm$^2$) and the myelin sheath radius distribution (mean=0.77 µm, variance=0.195 µm$^2$).

Following the approach described by [41], the effective radius can be approximated by the weighted-mean radius

$$a_{eff} \approx \frac{\int_0^\infty aP(a)N(a)\,da}{\int_0^\infty P(a)N(a)\,da}, \qquad (27)$$
$$= \frac{\int_0^\infty a^2P(a)\,da}{\int_0^\infty aP(a)\,da}$$

where $N(a)$ is the number of diffusing particles as a function of the radius $a$. In our case, $N(a)$ is proportional to the surface area of the cylinder and, therefore, to its radius. Consequently, the signal contribution from each cylinder is approximately proportional to its radius. Thus, we expect $a_{eff}$ to correlate with the ratio $\langle a^2 \rangle / \langle a \rangle$ determined from the underlying distribution $P(a)$.

Alternatively, another approximation can be obtained by following the approach presented by [37] using the Gaussian approximation with time-dependent radial diffusivity. When assuming small myelin sheath radii such that $D_{\|}t >> a^2$ and $D_{\|} >> D_{\perp}^{app}$, for low and moderate *b*-values, the normalized spherical mean dMRI signal can be approximated by:

$$\langle S(q,\Delta,\delta,\xi,a) \rangle \approx \sqrt{\frac{\pi}{4}} \frac{erf\left(\sqrt{bD_{\|}}\right)}{\sqrt{bD_{\|}}} \left(1 - b\frac{a^2}{2t_{\exp}}\right), \qquad (28)$$

where we used Eqs. (18), (13) and Lori's correction. Inserting this equation into the right-hand side of Eq. (26) and equating this expression to the signal arising from a single cylindrical surface with radius $a_{eff}$ we obtain:

$$\left\langle S\left(q,\Delta,\delta,\xi,a_{eff}\right)\right\rangle \approx \sqrt{\frac{\pi}{4}}\frac{erf\left(\sqrt{bD_{\parallel}}\right)}{\sqrt{bD_{\parallel}}}\frac{\int_0^\infty aP(a)\left(1-b\frac{a^2}{2t_{\exp}}\right)da}{\int_0^\infty aP(a)\,da},$$

$$=\sqrt{\frac{\pi}{4}}\frac{erf\left(\sqrt{bD_{\parallel}}\right)}{\sqrt{bD_{\parallel}}}\left(1-\frac{b}{2t_{\exp}}\frac{\int_0^\infty a^3P(a)\,da}{\int_0^\infty aP(a)\,da}\right). \quad (29)$$

Comparing Eqs. (29) and (28) we obtain

$$a_{eff}^2=\frac{\int_0^\infty a^3P(a)\,da}{\int_0^\infty aP(a)\,da}\quad . \quad (30)$$

Thus, we might also expect $a_{eff}$ to correlate with the expression $\left(\left\langle a^3\right\rangle/\left\langle a\right\rangle\right)^{1/2}$.

In the Results section, we will compare these two effective radius definitions with the numerical effective radius determined by fitting Eq. (26) to the theoretical model corresponding to a single cylinder. This evaluation will use histological measurements of inner axon radii sampled from four regions of the Corpus Callosum in a human brain [91], which will be converted into distributions of myelin sheath radii according to Eq. (25).

## 3. Methods

### 3.1 Monte Carlo simulations

Monte Carlo Diffusion Simulations (MCDS) were employed as a benchmark to validate the proposed models. We used an MC simulator developed by our group, available at https://github.com/jonhrafe/Robust-Monte-Carlo-Simulations [92]. This simulator has been validated against analytical models across multiple geometries, including impermeable planes, cylinders, and spheres [92]. For this study, we extended its capabilities to incorporate new myelin water diffusion models, implementing two geometrical structures: 3D infinite, impermeable cylinders and spiral surfaces.

The analytical models were validated by comparing their predicted dMRI signals to those generated by the MC simulations for identical impermeable cylindrical surfaces. Additionally, the dMRI signals from concentric cylinders were compared with those from spiral surfaces to assess the assumption presented in Section 2.1 (Figure 1). This assumption suggests that net radial displacements along the spiral trajectory are negligible, which allows the diffusion process in the more complex spiral geometry to be approximated as that in concentric cylinders.

## 3.2 Geometrical Models

### 3.2.1 Cylindrical Surfaces

We simulated diffusion on infinite, impermeable cylindrical surfaces. The diffusion process was simulated using a fixed step size along both the z-axis (aligned with the main axis of the cylinder) and the curved trajectory in the x-y plane, given by $l = \sqrt{2D_{\|} t / N_t}$ , where $N_t$ is the number of Monte Carlo steps and $t$ is the total diffusion time. At each step, the particle's z-coordinate was updated as $z \leftarrow z \pm l$ , with the direction randomly selected to simulate upward and downward motion. In the x-y plane, the angular displacement was selected to maintain a constant arc length $l$, i.e., $\theta \leftarrow \theta \pm l/a$, allowing particles to move in either rotational direction. The radius $a$ was constant, reflecting the cylindrical surface's geometry.

For each *b*-value, dMRI signals were generated from 50 independent cylinders with radii uniformly spaced from 0.1 μm to 5.0 μm in increments of 0.1 μm. To simulate the myelin water dMRI signal from a single axon with specific inner and outer radii, we calculated the radius-weighted sum of the signals from all cylindrical surfaces in this range, following Eq. (19).

To replicate the myelin water dMRI signal based on voxelwise realistic distributions of myelin radii, we performed the following steps:

1. Converted histological distributions of inner axon radii from [91] into myelin sheath radii using Eq. (25), assuming a constant g-ratio of 0.7.
2. Computed the spherical mean dMRI signal for each resulting distribution by evaluating the integral in Eq. (26), discretized using the same grid of 50 radii ranging from 0.1 to 5.0 μm as used in the MC simulations.

### 3.2.2 Spiral Surfaces

The diffusion process was similarly simulated for the spiral surfaces using a fixed step size $l$ along the z-axis and the x-y plane. The curved trajectory in the x-y plane was determined by the particle's position on the spiral. The radius $a(\theta)$ of the spiral varies with the polar angle $\theta$ in the x-y plane, according to $a(\theta) = a_i + (d_s/2\pi)\theta$, where $a_i$ is the inner radius and $d_s$ is the distance between successive layers of the spiral. The inter-layer distance was fixed to $d_s = d_m + d_w$ =7.5nm, based on histological data reported by [73]. In this context, $d_m$ and $d_w$ represent the thickness of the myelin layer and the spacing filled by myelin water, respectively. Therefore, $d_s$ corresponds to the distance between the centers of the gaps filled by myelin water in an axon. The polar angle $\theta$ ranged from 0 to the maximum value for which $a(\theta) = a_o$.

To assess whether the dMRI signals from water molecules confined to spiral surfaces can be approximated by those from concentric cylindrical surfaces, we generated spiral geometries with g-ratios of 0.6, 0.7, and 0.8, consistent with values reported in histological studies [93,94]. Since the results across different g-ratios were comparable, we present findings for g-ratio=0.7, using three geometries with inner and outer radii of 0.5/0.7 μm, 0.7/1.0 μm, and 1.0/1.4 μm, respectively.

The resulting signals were compared to those from corresponding cylindrical surfaces using the same PGSE sequence parameters described in the next section.

### 3.3 Simulation Protocol

The diffusion process was simulated for both geometrical models using a total diffusion time of $t$=20 ms and $N_t$=15,000 steps per particle. We conducted a bootstrap-based analysis to ensure convergence of the simulations, as outlined in [92]. A total of 75,000 particles were uniformly distributed on each cylindrical or spiral surface. Three values of parallel diffusivity ($D_{\parallel}$=0.3, 0.5, 0.8 μm$^2$/ms) were used to cover the range of myelin water diffusivities reported by [65].

A PGSE sequence with trapezoidal diffusion gradients was used to generate dMRI signals. The sequence was based on the specifications of a Connectome 2.0 scanner, employing a maximum gradient strength of *G*=500 mT/m and a maximum slew rate of *SR*=600 T/m/s [68], yielding to a trapezoidal ramp rise time $\xi = G/SR$ =0.833 ms. The protocol included 90° and 180° pulse durations of 2 ms and 4 ms, respectively. Six *b*-values were selected using the shortest possible TE for each case while maintaining maximum *G* and *SR*, following the implementation described in [70] and [71]. Table 1 details the experimental parameters.

For each *b*-value, dMRI signals were generated for 92 gradient orientations uniformly distributed on the unit sphere, along with the signal for $b$=0. The subsequent analyses focused on the spherical mean signal normalized by the $b$=0 signal.

Insert Table 1 around here.

**Table 1**. Experimental parameters for Monte Carlo simulations using a PGSE sequence with trapezoidal diffusion gradients. The simulations employed a diffusion gradient strength of $G$=500 mT/m and a slew rate of $SR$=600 T/m/s. For each experiment, 92 gradient orientations were uniformly distributed on the unit sphere.

| $b$ (ms/μm²) | $\Delta$ (ms) | $\delta$ (ms) | TE (ms) |
|---|---|---|---|
| 0.8 | 7.45 | 2.62 | 12.90 |
| 1.0 | 7.72 | 2.88 | 13.43 |
| 1.5 | 8.27 | 3.44 | 14.54 |
| 2.0 | 8.72 | 3.89 | 15.44 |
| 2.5 | 9.11 | 4.27 | 16.21 |
| 3.0 | 9.45 | 4.61 | 16.89 |

## 4. Results

### 4.1 Diffusion diffraction pattern: single cylinder

Figure 3 illustrates the theoretical spherical mean dMRI signal from a cylindrical surface, as generated by the general model presented in Eq. (17) using a PGSE sequence with trapezoidal diffusion gradients. The signal is shown for *b*-values ranging from 0 to 100 ms/μm$^2$ and for three cylinders with radii of 0.3 μm, 1.0 μm, and 3.0 μm.

For relatively low *b*-values (approximately below 3 ms/μm$^2$), the logarithm of the signal approximates a linear relationship. This linearity suggests that a Gaussian model could be valid in this regime. However, as the *b*-value increases, deviations from Gaussianity become apparent, and signal oscillations, known as diffraction patterns, emerge. These diffraction-like patterns have been reported in other geometries where diffusion is confined, such as planar, cylindrical, and spherical domains [95–97].

Insert Figure 3 around here (1.5 columns)

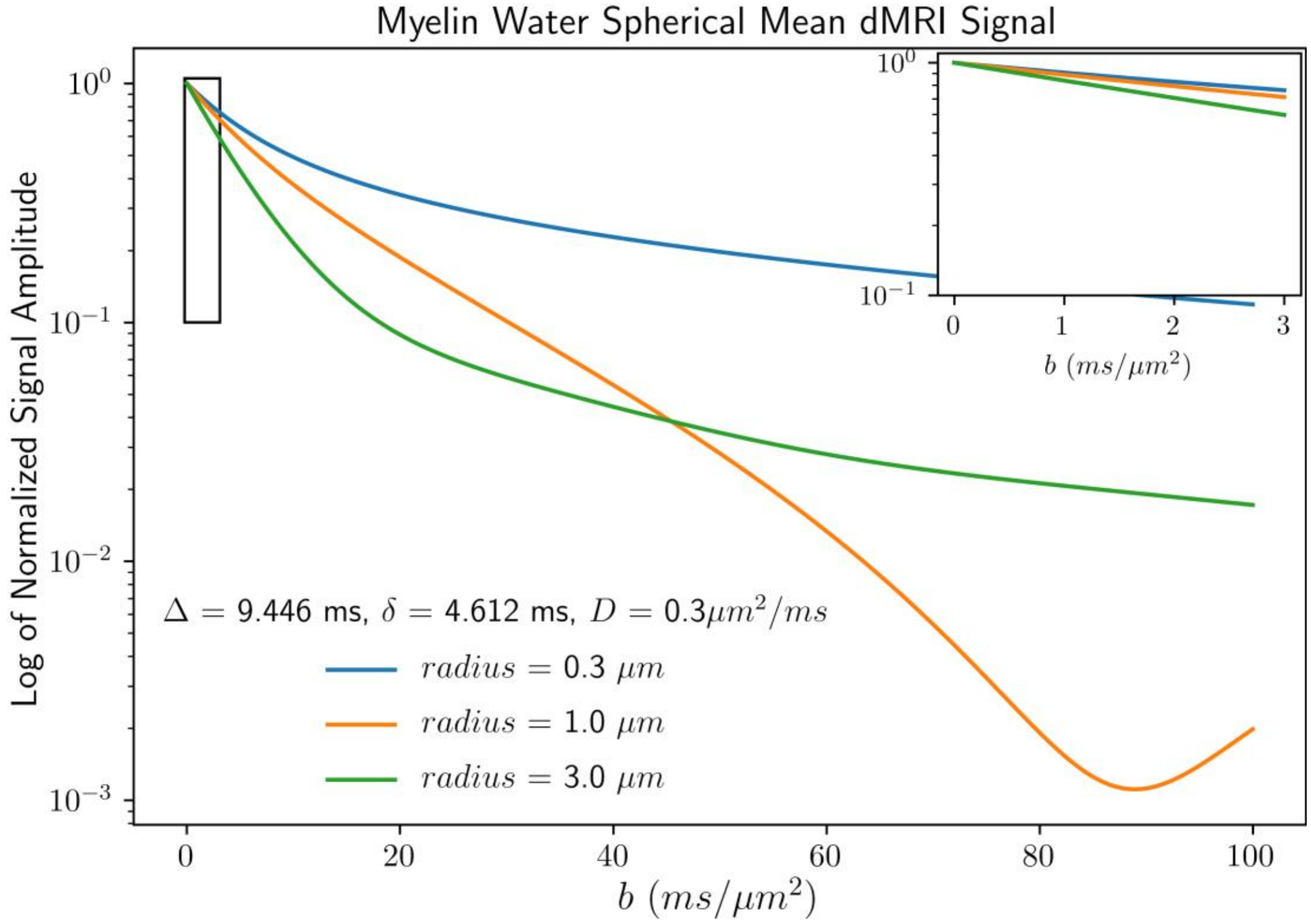

**Figure 3. Theoretical spherical mean signal attenuation for cylindrical surfaces.** The signal was generated using the general model presented in Eq. (17) for *b*-values ranging from 0 to 100 ms/μm², with diffusion time parameters of $\Delta = 9.446$ ms, $\delta = 4.612$ ms, and $D_{\|} = 0.8$ μm²/ms. The signal decay for *b*-values from 0 to 3 ms/μm² is displayed in a separate zoomed-in axis, as indicated by the rectangular box. The signal attenuation is plotted on a logarithmic scale for three cylinders with radii of 0.3 μm (blue), 1.0 μm (green), and 3.0 μm (orange) as a function of the *b*-value.

### 4.2 Single cylinder dMRI signal using 'realistic' acquisition parameters vs MC simulations

To assess the accuracy of the new analytical models proposed in this study, we compared the predicted dMRI signals with those generated by MC simulations. Figure 4 shows the theoretical spherical mean dMRI signals from cylindrical surfaces as a function of the radius, as predicted by both the general analytical model and the Gaussian approximation with time-dependent radial diffusivity (Eqs. (17) and (18), respectively) using a PGSE sequence with trapezoidal diffusion gradients. Additionally, the figure includes the dMRI signals obtained from the MC simulations for validation purposes. This comparison was conducted over a range of parallel diffusivities ($D_{\|}$=0.3, 0.5, 0.8 μm$^2$/ms) and practical *b*-values from 0.8 to 3.0 ms/µm², achievable in preclinical and human scanners equipped with strong diffusion gradients.

Increasing the *b*-value results in greater attenuation of the dMRI signal as a function of the radius across all three diffusivity values. At a *b*-value of 3.0 ms/µm², the signal exhibits maximum sensitivity to myelin sheath radii in the 0.5 to 3.0 µm range. However, at this higher *b*-value, we observe the largest, albeit still minor, deviations between the signals predicted by the analytical models and those generated by the MC simulations. Notably, the agreement between the models and simulations is strongest for the lowest diffusivity ($D_{\parallel}$=0.3 µm$^2$/ms, panel A). It diminishes as diffusivity increases, with the largest discrepancy observed at $D_{\parallel}$=0.8 µm$^2$/ms (panel C).

For this acquisition protocol, the signal shows minimal sensitivity to myelin radii smaller than 0.5 µm and larger than 3.5-4.0 µm. This result indicates that the method is best suited for detecting myelin sheath sizes in the 0.5-3.5 µm range. Across all *b*-values, the Gaussian approximation closely follows the analytical model, particularly for radii below 4.0 µm, further confirming the accuracy of the approximation in this parameter range.

Insert Figure 4 around here (1.25-1.5 columns)

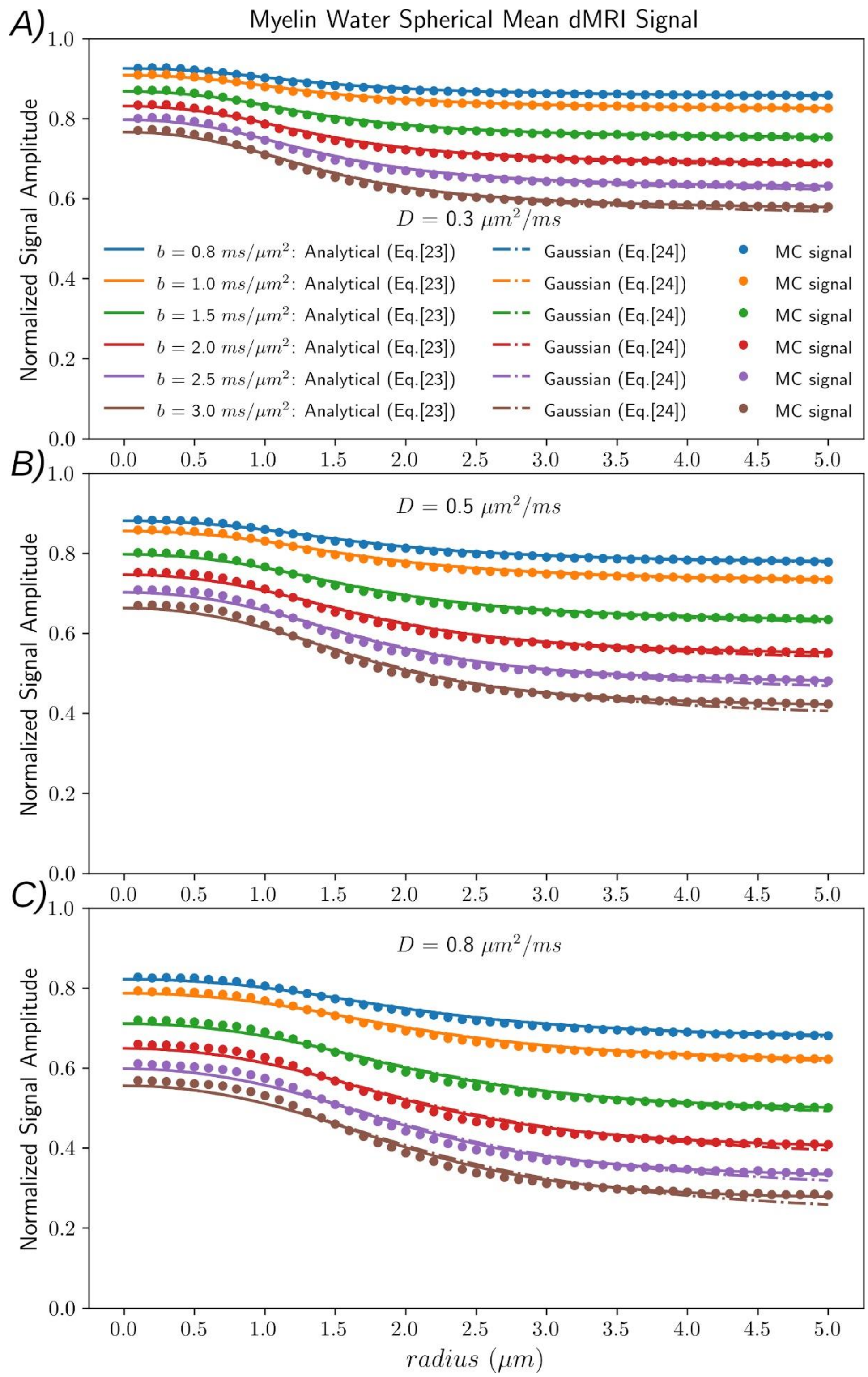

**Figure 4. Sensitivity of the spherical mean dMRI signal as a function of myelin sheath radii for different diffusivities**. The signals were generated using the general model (Eq. (17), continuous lines), the Gaussian approximation (Eq. (18), dashed lines), and Monte Carlo (MC) numerical simulations (dots) for the following *b*-values: [0.8, 1.0, 1.5, 2.0, 2.5, 3.0] ms/μm², using a PGSE sequence with parameters listed in Table 1. Panels A), B), and C) show results corresponding to diffusivities of $D_{\parallel} = 0.3$ , $D_{\parallel} = 0.5$ , and $D_{\parallel} = 0.8$ μm²/ms, respectively.

The normalized signal amplitudes from the analytical models are displayed for myelin sheath radii ranging from 0 to 5 µm, and the MC signals were generated for 50 discrete radii ranging from 0.1 to 5 µm.

### 4.3 Spiral surfaces vs concentric cylinders: MC simulations and analytical models

The results from the experiment comparing the spherical mean dMRI signals generated by MC simulations for spiral geometries and multiple concentric cylinders are presented in Figure 5. Specifically, Figure 5 shows the dMRI signals as a function of the six *b*-values employed. The signal from a spiral geometry with inner and outer radii of 0.7 µm and 1.0 µm is compared with the radius-weighted signal from multiple concentric cylinders within the same radius range, calculated using Eq. (19). Additionally, we display the signals from individual cylindrical surfaces with radii ranging between 0.7 µm and 1.0 µm, obtained from both MC simulations and the analytical models. Panels A and B correspond to results for diffusivities of $D_{\parallel}$=0.3 µm$^2$/ms and $D_{\parallel}$=0.8 µm$^2$/ms, respectively.

For both diffusivity values, we observe a strong agreement between the MC-generated signals for the spiral geometry and the radius-weighted aggregation of signals from concentric cylinders with the same range of radii. This result suggests that the spiral geometry can be accurately approximated by multiple concentric cylinders. Notably, for the lower diffusivity ($D_{\parallel}$=0.3 µm$^2$/ms, panel A), the analytical model's predictions for individual cylinders closely match the signals generated by MC simulations. Furthermore, the signal produced by the spiral geometry is very similar to that of a single cylinder with a radius intermediate to the inner and outer radii. This implies that when fitting these signals with a single-radius model, the estimated effective radius would likely correspond to a value close to the average radius of the spiral.

However, for simulations at the higher diffusivity ($D_{\parallel}$=0.8 µm$^2$/ms, panel B), the signal decay predicted by the analytical models as a function of the *b*-value is more pronounced than the decay observed in the MC simulations. This result indicates potential inaccuracies in the analytical model at higher diffusivities and larger *b*-values. Consequently, the effective radius predicted by the analytical models will likely be biased towards a smaller value than the actual radius.

The results for spirals with other inner and outer radii were consistent with these findings. Specifically, the observed discrepancy for $D_{\parallel}$=0.8 µm$^2$/ms was reduced for the spiral with a larger inner radius of 1.0 µm. Conversely, the disagreement increased for the smaller spiral with an inner radius of 0.5 µm (results not shown).

Insert Figure 5 around here (1.5 columns)

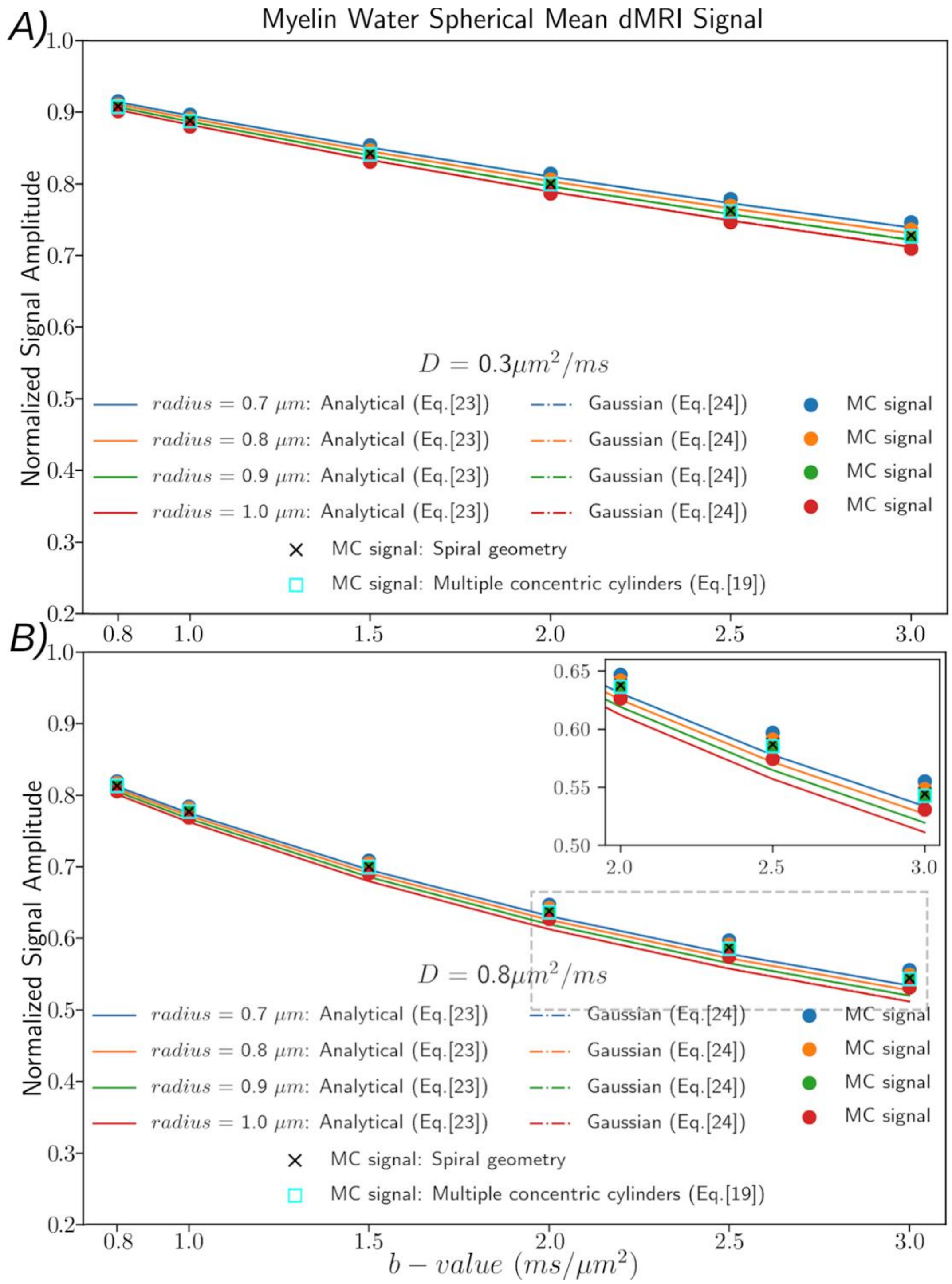

**Figure 5. Comparison of dMRI signals from spiral surfaces and concentric cylinders.** Spherical mean dMRI signals as a function of six employed *b*-values and results from Monte Carlo (MC) simulations for spiral geometries and multiple concentric cylinders. The signals are generated for a spiral with inner and outer radii of 0.7 µm and 1.0 µm, respectively, alongside radius-weighted signals from concentric cylinders within the same radius range. Signals from individual cylindrical surfaces with radii between 0.7 µm and 1.0 µm are plotted using both MC

simulations and analytical models. Panels A) and B) show results for $D_{\|} = 0.3$ µm²/ms and $D_{\|} = 0.8$ µm²/ms, respectively. In panel B), we highlight a region where the most significant discrepancies were observed between the signals computed using the analytical models and those obtained from the MC simulations.

### 4.4 Effective radius from histological measurements for distributions of cylinders

Figure 6 compares the effective radii estimated from simulated dMRI data against three different metrics derived from the distribution of myelin sheath radii in four regions of interest within the Corpus Callosum: axons connecting the prefrontal, motor, parietal, and visual cortices. The inner axon radii for these regions, as reported by [91], were modeled using Gamma distributions. These distributions were subsequently transformed into myelin sheath radii distributions using Eq. (25) and a constant g-ratio of 0.7.

We then generated the spherical mean dMRI signals corresponding to these distributions by discretizing Eq. (26) and employing the MC simulated signals. We assumed a parallel diffusivity of $D_{\|}$=0.5 µm$^2$/ms. The generated signals were fitted to the general single-cylinder model in Eq. (17) to estimate the effective radius. Figure 6 presents the effective radii $a_{eff}$, the mean radii $\langle a \rangle$ obtained from the distributions, and the second- and third-moment-based radii metrics $\langle a^2 \rangle / \langle a \rangle$ and $\left( \langle a^3 \rangle / \langle a \rangle \right)^{1/2}$, as defined in Eqs. (27) and (30).

The myelin sheath radii distributions in Figure 6 exhibit slightly longer right-hand tails and lower frequency values for small radii compared to the inner axon radii distributions, as expected. This difference arises because the myelin sheath radii represent all possible layer radii within the range defined by the inner and outer radii for all axons. Hence, it includes contributions from myelin layers near the inner and outer boundaries. These two distributions converge further as the g-ratio increases, as described by Eq. (25). This trend is noticeable when comparing the distributions in Figure 2 for a g-ratio of 0.6 with those in Figure 6 employing a g-ratio of 0.7.

The results show that for the distributions with smaller radii (Prefrontal and Parietal regions), the estimated effective radius $a_{eff}$ closely matches the mean radius $\langle a \rangle$. However, for the Motor and Visual regions, with larger radii distributions, the effective radius aligns more closely with the second-moment-based metric $\langle a^2 \rangle / \langle a \rangle$, followed by the third-moment-based metric $\left( \langle a^3 \rangle / \langle a \rangle \right)^{1/2}$. These findings suggest that the appropriate descriptor of the distribution may depend on the range of radii in each region.

Insert Figure 6 around here (2 columns)

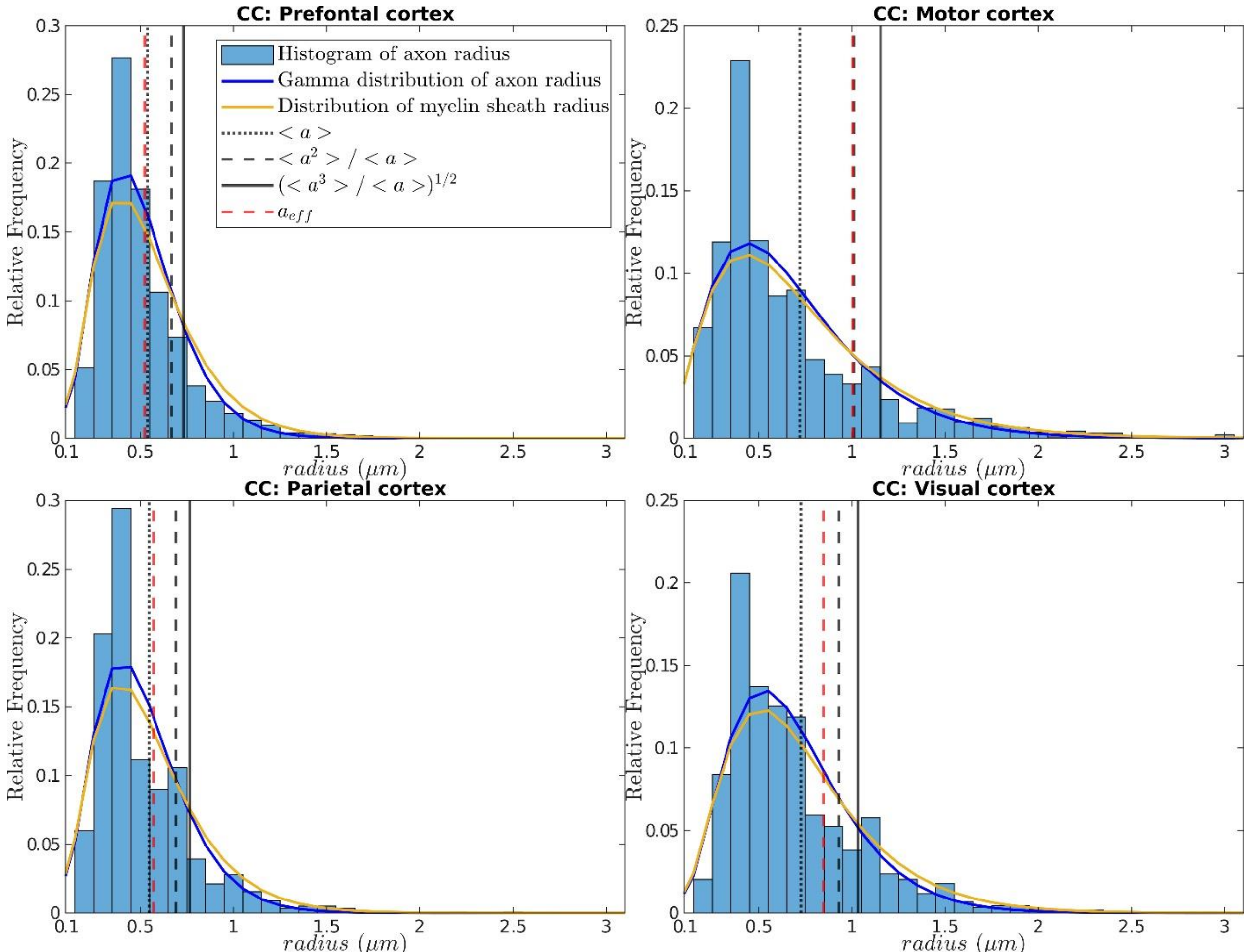

**Figure 6. Distributions of inner axon and myelin sheath radii and estimated effective radius**. Four subplots are presented, each corresponding to a different region of interest in the Corpus Callosum of a human brain. Each subplot includes a histogram of the measured inner axon radius (data from [91]), along with the best-fitting Gamma distribution (in blue) and the derived myelin sheath radius distribution estimated using Eq. (25) (in yellow-orange). The effective radius $a_{eff}$, estimated as the radius from the single-cylinder model (see Eq. (17)) that best fits the signal generated from the whole distribution of myelin sheath radius (see Eq. (26)), is plotted, along with three representative metrics of the distribution, including the mean value $\langle a \rangle$ and the second- and third-moment based metrics $\langle a^2 \rangle / \langle a \rangle$ and $\left( \langle a^3 \rangle / \langle a \rangle \right)^{1/2}$ derived in Eqs. (27) and (30), respectively. These results correspond to simulations using $D_{\parallel}$=0.5 $\mu m^2$/ms.

To further investigate the relationships between the effective radius and the derived metrics from the myelin sheath radii distributions, we present a correlation analysis in Figure 7. This figure illustrates the correlations between the effective radius and the three descriptive metrics across experiments conducted with three distinct diffusivities.

As shown in Figure 7, although these metrics reflect different aspects of the myelin sheath radii distributions, they exhibit significant correlations with the effective radius. Notably, the second-moment-based radius $\langle a^2 \rangle / \langle a \rangle$ demonstrated the strongest linear correlation (and smallest p-value) with $a_{eff}$ across all diffusivity values, indicating its potential as a reliable descriptor of effective radii. The third-moment-based radius $\left( \langle a^3 \rangle / \langle a \rangle \right)^{1/2}$ closely followed this trend, while the average radius showed less strong correlations. Interestingly, the analysis reveals a trend where the estimated effective radius tends to decrease with increasing diffusivity, particularly pronounced in distributions characterized by smaller axon radii.

Insert Figure 7 around here (2 columns)

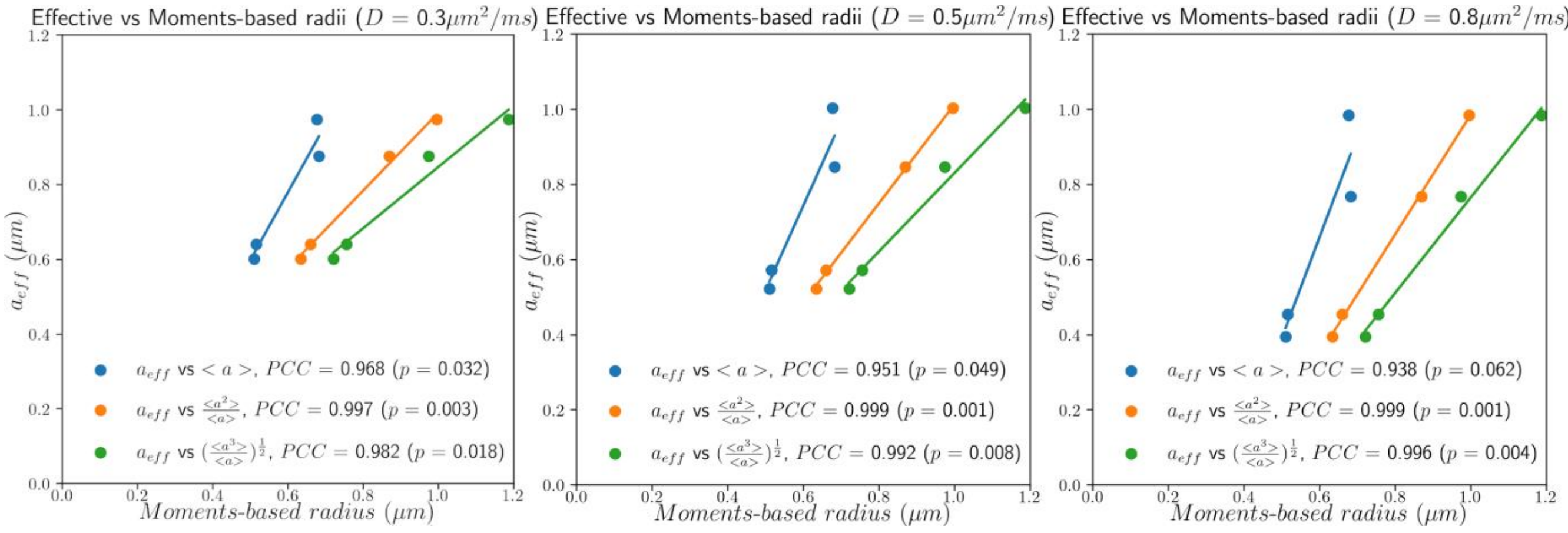

**Figure 7. Correlation between the effective radius and descriptive metrics.** This figure shows the correlations between the effective radius $a_{eff}$ (y-axis), estimated from dMRI signals, and three derived metrics (x-axis) from myelin sheath radii distributions: mean radius $\langle a \rangle$ (blue), second-moment-based radius $\langle a^2 \rangle / \langle a \rangle$ (orange), and third-moment-based radius $\left( \langle a^3 \rangle / \langle a \rangle \right)^{1/2}$ (green). Panels A, B, and C depict the results for three distinct diffusivity values: $D_{\parallel}$=0.3, 0.5, and 0.8 μm$^2$/ms. The Pearson's Correlation Coefficient (PCC) and the corresponding p-value are reported for each analysis. Each set of points represents the values estimated from the four distributions shown in Figure 6.

## 5. Discussion

In this proof-of-concept study, we developed two models for the dMRI signal arising due to water molecular displacements on cylindrical surfaces. We focused on potential applications for modeling the dMRI signal associated with myelin water in brain tissues. In the first, more general model, we derived an exact analytical expression for the dMRI signal using the diffusion propagator formalism based on the narrow pulse approximation. The second model employs a Gaussian approximation with time-dependent radial diffusivity, offering a simpler analytical relationship. We also developed approximate signal expressions for PGSE protocols with trapezoidal and rectangular diffusion gradients, extending beyond the narrow pulse assumption.

We derived the spherical mean signal expressions for both models, which are theoretically independent of axonal orientation effects. The spherical mean signal remains invariant to orientation dispersion, as it is approximately equivalent to whether the axons within a voxel have varying orientations or are aligned to the same orientation [33,98]. While it is theoretically feasible to estimate both fiber orientations and the effective radius of the myelin sheath, such fitting procedures may be unstable. To address this challenge, we adopted a strategy inspired by previous studies on axon diameter mapping. These studies also employ the spherical mean approach to minimize the influence of orientation effects [37,43,46,99], a well-known confounding factor that can bias axon diameter estimates. Indeed, when the dispersion is not accurately incorporated into the model, it could alter the estimated radial and parallel myelin water diffusivities. Conversely, when the spherical mean signal is used, the estimated diffusivities more accurately reflect the intrinsic diffusivities of myelin water.

We also derived expressions for the dMRI signal from multiple concentric cylinders as the radius-weighted sum of signals. This was done to account for the dependence of signal intensities on the cylinders' surface areas and, thus, their radii. We further generalized this approach to consider a distribution of myelin sheath radii. Various approximations were introduced to enhance our understanding of the effective radius—the radius estimated by fitting the signal from a radius distribution to a single-radius model. Finally, we extended our MC diffusion simulation toolbox to simulate the diffusion process confined on cylindrical and spiral surfaces to compare the analytical and numerical dMRI signals.

Validating the proposed models would require comparing the effective radii estimated from dMRI data and the corresponding values measured from histology on the same brain regions. However, since histological studies typically report only the inner radius distribution, we introduced a new analytical approach to convert this distribution into a distribution of myelin sheath radii based on the assumption of a constant g-ratio across all axons in the sample. It is important to emphasize that this analytical relationship is primarily a practical tool for leveraging existing histological data. If new histological studies provide direct measurements of myelin sheath radii, we would no longer need to rely on this approximation for validation.

The proposed models can potentially estimate the effective myelin sheath radius from real dMRI data. For example, our models could be directly applied in diffusion-T1 experiments using inversion recovery sequences that effectively isolate signals from myelin water, as outlined in [65]. Similarly, for acquisition sequences where signals from other compartments are not entirely suppressed—such as in diffusion-T2 hybrid sequences proposed by [63] and [64] or the magnetization-prepared dMRI sequence described by [66]—our models could be integrated with existing multi-compartment dMRI frameworks, e.g., [31,33,100,101], to concurrently fit the myelin water component along with parameters for other compartments. Additional investigations are needed to identify the optimal acquisition protocols for these multi-compartment fittings, focused on mitigating model fitting degeneracies [102]. These approaches could be applied to both ex vivo and in vivo data using scanners with strong diffusion gradients, leveraging recent advances [70–72] that enhance the myelin water dMRI signal by reducing echo times.

Our MC simulations employed parallel diffusivity values as reported by [65], specifically $D_{\parallel}$=0.37 μm$^2$/ms in excised frog sciatic nerve for the double-inversion-recovery sequence. Since their experiments were conducted within one hour post-euthanasia and lasted approximately 90 minutes, this relatively short post-mortem interval likely helped preserve some of the tissue's original diffusion properties compared to in vivo studies, thereby minimizing significant alterations due to dehydration or tissue degradation. However, the reduced temperature (20°C) relative to the typical in vivo temperature (around 37°C) may have contributed to decreased diffusivity. Hence, we expect the diffusivity values they reported to be lower than those observed in vivo. On the other hand, we anticipate that myelin water diffusivity will be lower than in other WM compartments due to its higher bound water content, which results in shorter relaxation times and reduced mobility. Therefore, we employed myelin water parallel diffusivities in the 0.3-0.8 μm$^2$/ms range.

This study is not the first to simulate the dMRI signal from myelin water. To our knowledge, two previous works have specifically addressed the multi-wrapping nature of myelin [103,104]. In the first study [103], this aspect was modeled implicitly by assuming a higher myelin water diffusivity in the tangential direction than the radial one. MC simulations were employed to assess the sensitivity of dMRI models to the diffusive properties of myelin water. Their findings indicate that myelin water could influence the apparent diffusion coefficient and kurtosis measured transverse to the orientation of WM tracts. In contrast, the second study [104] conducted MC simulations to examine water exchange through myelin sheaths by explicitly creating a spiraling myelin structure. They observed sub-second exchange times for thin axons with fewer wraps, highlighting the importance of modeling water exchange across WM compartments, especially in clinical studies on demyelinating diseases and the developing infant brain. Conversely, a slow exchange rate was observed in axons with more than eight myelin sheaths, typical of healthy WM in humans, supporting the assumption of impermeable membranes.

While other methods exist for quantifying WM microstructure parameters, including the inner axon radius and myelin content, each has inherent limitations. Myelin volume, often combined with the fiber volume fraction estimated from dMRI data to calculate the mean g-ratio, is typically determined using Magnetization Transfer (MT) or Multi-echo T2 (MET2) relaxometry techniques. However, although MT and MET2 techniques are known for their sensitivity to changes in myelin content, they are not exclusively specific to myelin, as other tissue compartments can also influence the measured signal [53,105,106]. Similarly, inner axon radius mapping techniques based on dMRI data face a resolution limit below which the radii of smaller axons cannot be reliably estimated [39,74]. As such, the estimated effective inner radius typically represents the right-hand tail of the inner axon radius distribution rather than the entire distribution [37]. As myelin imaging techniques (i.e., MT and MET2) are not affected by the same resolution limit, care should be taken when combining estimates from these techniques to predict total myelin thickness (i.e., the difference between the outer and inner axon radii).

To the best of our knowledge, we present the first models for estimating myelin sheath radii exclusively using dMRI data, offering a novel imaging biomarker for detecting changes in myelin thickness. Although the method does not directly estimate the distance between the inner and outer layers of the myelin, it provides an integrated measure representing the entire distribution of myelin layer radii. The effective myelin sheath radius is derived by fitting a single-cylinder-surface model to the dMRI signal. In a hypothetical sample of identical axons with the same g-ratio, the effective radius closely approximates the mean of the inner and outer axon radii. In more realistic scenarios, where axon radii vary, and each axon has a distinct g-ratio, it reflects a population-weighted average with larger myelin layers contributing more substantially to the overall value.

Although our results are promising, several limitations need to be addressed in future work:

i. While the analytical models closely match MC simulations under various experimental conditions, discrepancies emerge at high *b*-values and large diffusivities. These inaccuracies arise from the approximations introduced to facilitate modeling. We initially derived our models using the narrow pulse approximation and later applied a correction framework to extend their applicability beyond this scheme. However, it is important to note that this correction framework primarily provides a valid approximation for Gaussian diffusion. The diffusion process deviates from Gaussian behavior in scenarios involving small cylinder radii, high diffusivities, and high *b*-values. One potential approach to address this limitation is to adapt the multiple propagator method introduced by [107] and refined by [108] to our specific models. Additionally, exploring a data-fitting approach based on a dictionary of precomputed MC signals may allow us to circumvent the limitations imposed by the theoretical approximations.

ii. The myelin sheath radius estimations are constrained by a resolution limit, influenced by both the strength of the diffusion gradient and the signal-to-noise ratio (SNR). For the employed acquisition parameters (i.e., $G_{max}$ = 500 mT/m), our results indicate that signals for myelin sheath radii smaller than 0.5 μm and higher than 3.5 μm are indistinguishable (Figure 4). This range shifts with the diffusion gradient strength: weaker gradients make it harder to detect smaller myelin sheaths, whereas stronger gradients, like those in preclinical scanners (e.g., $G_{max}$ = 1500 mT/m), improve sensitivity to thinner layers. We did not conduct a formal resolution analysis akin to [39,74] for estimating inner axon diameters, which would involve determining the exact resolution limit and its dependence on $G_{max}$ and SNR. However, combining measurements acquired with different diffusion gradient strengths could extend the sensitivity range, although this approach is more feasible in preclinical settings where stronger diffusion gradients are available. In practice, the myelin water dMRI signal attenuation is primarily influenced by myelin layers with radii within the detectable range, with greater sensitivity to the right-hand tail of the radii distribution. Therefore, clinical applications should target pathologies involving larger axons, as smaller myelin layers may fall below the resolution limit. This limitation is not unique to our method. Similar constraints affect other dMRI-based techniques, such as those used to estimate inner axon diameters [37,109].

iii. This study does not include a numerical evaluation of the model's robustness to noise and artifacts in dMRI data. The numerical stability depends on the specific dMRI sequence and experimental parameters, such as diffusion gradient strength, diffusion times, and TE. For example, combining diffusion-weighted and double-inversion recovery sequences optimized to suppress non-myelin water signals would enable direct fitting of the proposed models to the measured data. In contrast, diffusion-T2 acquisitions require a multi-compartment model incorporating the proposed methodology. In future work, we plan to address these issues, employing Cramér-Rao bound analyses to optimize acquisition parameters for different sequences and evaluate the fitting stability under varying noise levels.

iv. All results presented in this study are based on synthetic signals derived from the proposed analytical models or MC simulations. Validation with real dMRI data, including histological analyses of various brain regions, is crucial for future work. Additionally, the diffusivity values used in this study are based on those reported by [65]. Still, variations in reported myelin water diffusivities in other experimental [63] and numerical studies [110–112] suggest the need for further work to reconcile these discrepancies and identify more accurate ex vivo and in vivo myelin water diffusivities.

v. Our MC simulations and proposed models assume straight cylinders, thus neglecting axonal undulations and beading, which are known to influence diffusion in WM [99,113–115]. Incorporating more realistic axonal geometries constitutes a critical direction for future research, as modeling these effects could enhance the generalizability of our approach. To address these limitations, we plan to conduct numerical evaluations to assess their impact on the estimated effective myelin radius and adapt the models to include geometrical variations informed by histological data. Axonal undulations and beading are expected to reduce the apparent parallel diffusivity and increase the radial diffusivity of myelin water relative to values observed for straight cylinders. Based on the relationship between the radius and myelin water diffusivities provided by the Gaussian approximation in Eq. (13), these effects would likely lead to overestimating the effective myelin radius compared to the actual value.

vi. In severe pathological conditions, such as certain multiple sclerosis lesions, where the myelin sheath breaks down and undergoes vacuolization, leading to the separation of adjacent spirals as well as axonal dissociation and degeneration [116,117], the assumptions underlying the proposed model are no longer valid. In such cases, increased water permeability and alterations in myelin water layer thickness would compromise the applicability of the proposed formalism. Therefore, this model is likely more suited for studying healthy brains and pathological conditions at earlier stages with milder alterations.

vii. All data were generated based on an acquisition protocol potentially feasible with a Connectome 2.0-like human scanner equipped with a diffusion gradient of 500 mT/m, where the TE can be further reduced by employing an image readout technique starting at the center of k-space (e.g., spiral). Future studies should investigate a range of acquisition protocols, including stronger diffusion gradients available in preclinical scanners [37], as well as the 300 mT/m diffusion gradients utilized in the Connectome 1.0 [67,109] and GE SIGNA MAGNUS scanners. The recently introduced MAGNETOM Cima.X clinical scanner, with a diffusion gradient strength of 200 mT/m, should also be considered. Determining the optimal acquisition parameters for each scenario is crucial for improving sensitivity to myelin sheath radii.

In summary, this work introduces dMRI models capable of characterizing myelin water diffusion, enabling the estimation of the effective myelin sheath radius per voxel. This water pool has been largely overlooked in previous dMRI studies due to the strong signal suppression it experiences when long TEs are used in clinical applications due to its short T2 relaxation time. However, recent advancements in dMRI sequences and the advent of MRI scanners equipped with stronger diffusion gradients make it possible to acquire dMRI signals significantly weighted by myelin water. This progress underscores the importance of having available models for this specific tissue compartment.

Nevertheless, the applicability of the proposed methodology is limited by hardware availability. Its use is restricted to a few human scanners with strong diffusion gradients and preclinical animal scanners with higher gradient strengths (e.g., $G$ = 300–1500 mT/m). This limitation highlights the need for broader access to such advanced MRI systems to fully exploit the potential of these models for both research and clinical applications. Additionally, pathologies involving vacuolization of myelin sheaths or significant separation of adjacent spirals result in altered myelin water layer thickness and increased permeability, which could compromise the validity of the proposed formalism. Consequently, the model is best suited for studies of healthy brains and pathological conditions at earlier stages, where tissue alterations are less severe.

By addressing the discussed limitations and validating the models with real dMRI data and histological measurements, future research may enhance the accuracy and applicability of the proposed models, contributing to the development of novel MRI biomarkers of WM tissue microstructure.

## 6. Acknowledgements and funding

EC-R was supported by the Swiss National Science Foundation (SNSF), Ambizione fellowship PZ00P2_185814 and SNSF grant 10000706. CMWT is supported by a Sir Henry Wellcome Fellowship (215944/Z/19/Z), and EF-G is supported by the SNSF, grant number 10000706. To facilitate open access, the author has applied a CC BY public copyright license to any Author Accepted Manuscript arising from this submission.

## 7. Author Contribution

EC-R: Conceptualization, formal analysis, data curation, funding, investigation, methodology, software, supervision, validation, visualization, interpretation of data, writing – original draft, writing – review & editing. CMWT: Investigation, methodology, supervision, software, validation, visualization, interpretation of data, writing – original draft, writing – review & editing. EF-G, DKJ, J-PT: Investigation, interpretation of data, writing – review & editing. JR-P: Data curation, formal analysis, investigation, methodology, supervision, software, validation, visualization, interpretation of data, writing – original draft, writing – review & editing. All authors contributed to the article and approved the submitted version.

## 8. Conflict of interest

The authors declare that the research was conducted without any commercial or financial relationships that could be construed as a potential conflict of interest.

## 9. Code and data availability statement

All the synthetic datasets generated in this study, the scripts implementing the proposed models, and the Figures presented in this article will be freely available at https://github.com/ejcanalesr/myelin-water-diffusion-models/.

## 10. Appendices

### Appendix A: Derivation of the radial diffusion signal

To solve Eq. (6), we define the initial spin distribution $P(\rho',\theta')$ as a uniform probability distribution on a circle/cylinder with radius $a$:

$$P(\rho',\theta') = \frac{1}{2\pi a}\delta(\rho'-a), \tag{31}$$

where $\delta(x)$ is a Dirac delta function: it is 1 for $x=0$ and 0 otherwise. Substituting Eq. (31) into Eq. (6) and integrating over $d\rho'$, we obtain

$$\frac{E_\perp(\mathbf{q}_{xy},t)}{E_\perp(\mathbf{q}_{xy}=0,t)} = \frac{1}{2\pi}\int_0^\infty\int_0^{2\pi}\int_0^{2\pi} P(\rho,\theta|,a,\theta',t)e^{iq_{xy}\rho\cos(\varphi+\theta)}e^{-iq_{xy}a\cos(\varphi+\theta')}\rho d\rho d\theta d\theta'. \tag{32}$$

Since the displacement of the particles is confined to the circle's circumference, the probability $P(\rho,\theta|,a,\theta',t) = P(\theta|\theta',a,t)P(\rho|a,t)$ can be written as the product of the normalized angular distribution $P(\theta|\theta',a,t)$ for moving from angle $\theta'$ to $\theta$ in time $t$ on the circle with radius $a$, and a delta function prohibiting any movement in the radial coordinate $P(\rho|a,t) = \delta(\rho-a)/a$ (which guarantees that $\int_{-\infty}^{\infty} P(\rho|a,t)\rho d\rho = 1$) that is appropriate for impermeable cylinders.

After plugging these equations into Eq. (32), and integrating over $d\rho$ we obtain,

$$\begin{aligned}\frac{E_\perp(\mathbf{q}_{xy},t)}{E_\perp(\mathbf{q}_{xy}=0,t)} &= \frac{1}{2\pi}\int_0^{2\pi}\int_0^{2\pi} P(\theta|\theta',a,t)e^{iq_{xy}a\cos(\varphi+\theta)}e^{-iq_{xy}a\cos(\varphi+\theta')}d\theta d\theta', \\ &= \frac{1}{2\pi}\int_0^{2\pi}\int_0^{2\pi} P(\Phi|,a,t)e^{iq_{xy}a\cos(\psi)}e^{-iq_{xy}a\cos(\psi-\Phi)}d\psi d\Phi,\end{aligned} \tag{33}$$

where we used the change of variables $\psi=\varphi+\theta$ and $\Phi=\theta-\theta'$ in the second equation.

Substituting Eq. (8) into Eq. (33), and using the following Jacobi-Anger expansions [118]

$$e^{iq_{xy}a\cos(\psi)} = J_0\left(aq_{xy}\right) + 2\sum_{n=1}^{\infty} i^n J_n\left(aq_{xy}\right)\cos\left(n\psi\right),$$

$$e^{-iq_{xy}a\cos(\psi-\Phi)} = J_0\left(aq_{xy}\right) + 2\sum_{m=1}^{\infty}\left(-i\right)^m J_m\left(aq_{xy}\right)\cos\left(m\left(\psi-\Phi\right)\right), \tag{34}$$

we obtain

$$\begin{aligned}
\frac{E_\perp\left(\mathbf{q}_{xy},t\right)}{E_\perp\left(\mathbf{q}_{xy}=0,t\right)} &= \frac{1}{\left(2\pi\right)^2}\int_0^{2\pi}\int_0^{2\pi}\left[1+2\sum_{p=1}^{\infty}e^{-p^2\frac{Dt}{a^2}}\cos\left(p\Phi\right)\right]\times \\
&\left[J_0\left(aq_{xy}\right)+2\sum_{m=1}^{\infty}\left(-i\right)^m J_m\left(aq_{xy}\right)\cos\left(m\left(\psi-\Phi\right)\right)\right]\times \\
&\left[J_0\left(aq_{xy}\right)+2\sum_{n=1}^{\infty}i^n J_n\left(aq_{xy}\right)\cos\left(n\psi\right)\right]d\psi\, d\Phi, \\
&= \frac{1}{\left(2\pi\right)^2}\int_0^{2\pi}\int_0^{2\pi}\Big[J_0^2\left(aq_{xy}\right)+2J_0\left(aq_{xy}\right)\sum_{n=1}^{\infty}i^n J_n\left(aq_{xy}\right)\cos\left(n\psi\right)+ \\
&2J_0\left(aq_{xy}\right)\sum_{m=1}^{\infty}\left(-i\right)^m J_m\left(aq_{xy}\right)\cos\left(m\left(\psi-\Phi\right)\right)+ \\
&4\sum_{n=1}^{\infty}\sum_{m=1}^{\infty}i^n\left(-i\right)^m J_m\left(aq_{xy}\right)J_n\left(aq_{xy}\right)\cos\left(m\left(\psi-\Phi\right)\right)\cos\left(n\psi\right)\Big]\times \\
&\left[1+2\sum_{p=1}^{\infty}e^{-p^2\frac{Dt}{a^2}}\cos\left(p\Phi\right)\right]d\psi\, d\Phi
\end{aligned} \tag{35}$$

It is convenient to integrate over $d\psi$ ,

$$\begin{aligned}
\frac{E_\perp\left(\mathbf{q}_{xy},t\right)}{E_\perp\left(\mathbf{q}_{xy}=0,t\right)} &= \frac{1}{\left(2\pi\right)^2}\int_0^{2\pi}\Bigg\{\int_0^{2\pi}J_0^2\left(aq_{xy}\right)d\psi+2J_0\left(aq_{xy}\right)\sum_{n=1}^{\infty}i^n J_n\left(aq_{xy}\right)\int_0^{2\pi}\cos\left(n\psi\right)d\psi+ \\
&2J_0\left(aq_{xy}\right)\sum_{m=1}^{\infty}\left(-i\right)^m J_m\left(aq_{xy}\right)\int_0^{2\pi}\cos\left(m\left(\psi-\Phi\right)\right)d\psi+ \\
&4\sum_{n=1}^{\infty}\sum_{m=1}^{\infty}i^n\left(-i\right)^m J_m\left(aq_{xy}\right)J_n\left(aq_{xy}\right)\int_0^{2\pi}\cos\left(m\left(\psi-\Phi\right)\right)\cos\left(n\psi\right)d\psi\Bigg]\times \\
&\left[1+2\sum_{p=1}^{\infty}e^{-p^2\frac{Dt}{a^2}}\cos\left(p\Phi\right)\right]\Bigg\}d\Phi, \\
&= \frac{1}{2\pi}\int_0^{2\pi}\left[J_0^2\left(aq_{xy}\right)+2\sum_{n=1}^{\infty}J_n^2\left(aq_{xy}\right)\cos\left(n\Phi\right)\right]\times \\
&\left[1+2\sum_{p=1}^{\infty}e^{-p^2\frac{Dt}{a^2}}\cos\left(p\Phi\right)\right]\Bigg\}d\Phi,
\end{aligned} \tag{36}$$

where we used the following identities,

$$\int_0^{2\pi} \cos\left(m\left(\psi-\Phi\right)\right) d\psi = \frac{\sin\left(m\Phi\right)-\sin\left(m\left(\Phi-2\pi\right)\right)}{m} = 0,$$
$$\int_0^{2\pi} \cos\left(n\psi\right) d\psi = \frac{\sin\left(2\pi n\right)}{n} = 0, \tag{37}$$
$$\int_0^{2\pi} \cos\left(n\psi\right)\cos\left(m\left(\psi-\Phi\right)\right) d\psi = \pi\cos\left(n\Phi\right)\delta\left(n-m\right).$$

The second and third identities are also helpful in integrating Eq. (36) over $d\Phi$

$$\begin{aligned}\frac{E_\perp\left(\mathbf{q}_{xy},t\right)}{E_\perp\left(\mathbf{q}_{xy}=0,t\right)} &= \frac{1}{2\pi}\Bigg[ J_0^2\left(aq_{xy}\right)\int_0^{2\pi} d\Phi + 2\sum_{n=1}^{\infty} J_n^2\left(aq_{xy}\right)\int_0^{2\pi}\cos\left(n\Phi\right)d\Phi + \\ &\quad 2J_0^2\left(aq_{xy}\right)\sum_{p=1}^{\infty} e^{-p^2\frac{Dt}{a^2}}\int_0^{2\pi}\cos\left(p\Phi\right)d\Phi \\ &\quad +4\sum_{n=1}^{\infty}\sum_{p=1}^{\infty} J_n^2\left(aq_{xy}\right)e^{-p^2\frac{Dt}{a^2}}\int_0^{2\pi}\cos\left(p\Phi\right)\cos\left(n\Phi\right)d\Phi\Bigg], \\ &= J_0^2\left(aq_{xy}\right)+2\sum_{p=1}^{\infty} J_p^2\left(aq_{xy}\right)e^{-p^2\frac{Dt}{a^2}}.\end{aligned} \tag{38}$$

**Appendix B: Derivation of the effective radial diffusivity**

In the following equation, we first represent the mean squared displacement in polar coordinates and later use Eq. (8) to compute the mean squared displacement on the circle, where we used the second and third identities reported in Eq. (37):

$$
\begin{aligned}
\left\langle \left| \mathbf{r}_{xy} \right|^2 \right\rangle &= \left\langle r_x^2 + r_y^2 \right\rangle, \\
&= \left\langle a^2 \left( \cos(\theta) - \cos(\theta') \right)^2 + a^2 \left( \sin(\theta) - \sin(\theta') \right)^2 \right\rangle, \\
&= \left\langle 2a^2 \left( 1 - \cos(\Phi) \right) \right\rangle, \\
&= \int_0^{2\pi} 2a^2 \left( 1 - \cos(\Phi) \right) P\left( \Phi |, a, t \right) d\Phi, \\
&= 2a^2 \left( 1 - \int_0^{2\pi} P\left( \Phi |, a, t \right) \cos(\Phi) \, d\Phi \right), \\
&= 2a^2 \left( 1 - \frac{1}{2\pi} \int_0^{2\pi} \left[ \cos(\Phi) + 2 \sum_{p=1}^{\infty} e^{-p^2 \frac{Dt}{a^2}} \cos(p\Phi) \cos(\Phi) \right] d\Phi \right), \\
&= 2a^2 \left( 1 - e^{-\frac{Dt}{a^2}} \right).
\end{aligned} \tag{39}
$$

## Appendix C: Derivation of Lori's correction approach

The precise expression for $\left\langle e^{i\phi} \middle| \mathbf{r} \right\rangle_{\Delta,\delta,\mathbf{g}}$ in Eq. (15) is generally unknown, making it difficult to estimate a closed-form analytical expression for $\left\langle e^{i\phi} \right\rangle_{\Delta,\delta,\mathbf{g}}$ from the diffusion propagator. Fortunately, it can be derived for some particular cases.

It is well known that for a Gaussian anisotropic diffusion process characterized by a diffusion tensor $\mathbf{D}$, the PGSE signal attenuation for a rectangular diffusion gradient is given by [76]

$$
\frac{\left\langle e^{i\phi} \right\rangle_{\Delta,\delta,\mathbf{g}}}{\left\langle e^{i\phi} \right\rangle_{\Delta,\delta,\mathbf{g}=0}} = e^{-\mathbf{q}^T \mathbf{D} \mathbf{q} [\Delta - \delta/3]} \quad . \tag{40}
$$

By plugging Eq. (40) into Eq. (15), and substituting the corresponding Gaussian anisotropic distribution of displacements producing such a signal, we note that the following equality must hold:

$$
e^{-\mathbf{q}^T \mathbf{D} \mathbf{q} [\Delta - \delta/3]} = \int_{\mathbb{R}^3} \frac{1}{\sqrt{(2\pi)^3 \left| 2\mathbf{D}(\Delta + \delta) \right|}} e^{-\frac{1}{2} \mathbf{r}^T [2\mathbf{D}(\Delta+\delta)]^{-1} \mathbf{r}} \left\langle e^{i\phi} \middle| \mathbf{r} \right\rangle_{\Delta,\delta,\mathbf{g}} d\mathbf{r} \, . \tag{41}
$$

Notably, $\left\langle e^{i\phi} \middle| \mathbf{r} \right\rangle_{\Delta,\delta,\mathbf{g}}$ can be determined via inverse induction from this relationship. First, let us rewrite Eq. (40) as

$$e^{-\mathbf{q}^T\mathbf{D}\mathbf{q}[\Delta-\delta/3]} = e^{-\frac{1}{2}\mathbf{q}'^T[2\mathbf{D}(\Delta+\delta)]\mathbf{q}'}, \tag{42}$$

where a scaled q-space vector $\mathbf{q}' = \mathbf{q}\sqrt{\dfrac{\Delta-\delta/3}{\Delta+\delta}}$ was introduced. Next, by applying the Fourier integral theorem to Eq. (42) - stating that if we take the inverse Fourier transform of a function and subsequently take the Fourier transform of the resulting expression, we retrieve the original function - we get:

$$\begin{aligned} e^{-\frac{1}{2}\mathbf{q}'^T[2\mathbf{D}(\Delta+\delta)]\mathbf{q}'} &= \int_{\mathbb{R}^3}\left(\int_{\mathbb{R}^3} e^{-\frac{1}{2}\mathbf{q}'^T[2\mathbf{D}(\Delta+\delta)]\mathbf{q}'} e^{-i\mathbf{q}'\mathbf{r}} d\mathbf{q}'\right) e^{i\mathbf{q}'\mathbf{r}} d\mathbf{r}, \\ &= \int_{\mathbb{R}^3} \frac{1}{\sqrt{(2\pi)^3 |2\mathbf{D}(\Delta+\delta)|}} e^{-\frac{1}{2}\mathbf{r}^T[2\mathbf{D}(\Delta+\delta)]^{-1}\mathbf{r}} e^{i\mathbf{q}'\mathbf{r}} d\mathbf{r}. \end{aligned} \tag{43}$$

By comparing Eqs. (41), (42) and (43) we obtain that $\left\langle e^{i\phi} \middle| \mathbf{r} \right\rangle_{\Delta,\delta,\mathbf{g}} = e^{i\mathbf{q}'\mathbf{r}}$ is a complex exponential similar to that in the q-space formalism but with the scaled q-vector $\mathbf{q}'$. Substituting this result into Eq. (15), we obtain the following approximation:

$$\frac{\left\langle e^{i\phi} \right\rangle_{\Delta,\delta,\mathbf{g}}}{\left\langle e^{i\phi} \right\rangle_{\Delta,\delta,\mathbf{g}=0}} \approx \int_{\mathbb{R}^3} P(\mathbf{r},\Delta+\delta) e^{i\sqrt{\frac{\Delta-\delta/3}{\Delta+\delta}}\mathbf{q}\mathbf{r}} d\mathbf{r}. \tag{44}$$

A similar result can be obtained for PGSE sequences with trapezoidal gradients by replacing the total diffusion encoding time $t_{\exp} = \Delta+\delta$ with $t_{\exp} = \Delta+\delta+\xi$, and the effective diffusion time $t_{eff} = \Delta-\delta/3$ with $t_{eff} = \Delta-\delta/3+\xi^3/30\delta^2-\xi^2/6\delta$.

**Appendix D: Derivation of the spherical mean signal**

Since the spherical mean signal is rotationally invariant, its value for a distribution of identical cylinders with arbitrary orientations is equal to that from cylinders oriented along the z-axis. The spherical mean of Eq. (11) is

$$\begin{aligned} \langle S \rangle &= \frac{1}{4\pi}\int_0^{2\pi}\int_0^{\pi} E(\mathbf{q},t)\sin(\beta)\,d\beta d\phi, \\ &= \frac{E(\mathbf{q}=0,t)}{4\pi}\int_0^{2\pi}\int_0^{\pi} e^{-q^2\cos(\beta)^2 D_{\|}t}\left[J_0^2(aq\sin(\beta)) + 2\sum_{p=1}^{\infty} J_p^2(aq\sin(\beta)) e^{-p^2\frac{D_{\|}t}{a^2}}\right]\sin(\beta)\,d\beta d\phi, \quad (45) \\ &= \frac{E(\mathbf{q}=0,t)}{2}\int_0^{\pi} e^{-q^2\cos(\beta)^2 D_{\|}t}\left[J_0^2(aq\sin(\beta)) + 2\sum_{p=1}^{\infty} J_p^2(aq\sin(\beta)) e^{-p^2\frac{D_{\|}t}{a^2}}\right]\sin(\beta)\,d\beta, \end{aligned}$$

where the integral by angle $\phi$ was computed straightforwardly since the signal is antipodal symmetric (i.e., constant) for all angles $\phi$.

The previous equation can be rewritten as

$$\langle S\rangle = \frac{E(\mathbf{q}=0,t)}{2}\left[I(0)+2\sum_{p=1}^{\infty} e^{-p^2\frac{D_{\|}t}{a^2}} I(p)\right], \tag{46}$$

where,

$$\begin{aligned} I(p) &= \int_0^{\pi} e^{-q^2\cos(\beta)^2 D_{\|}t} J_p^2\left(aq\sin(\beta)\right)\sin(\beta)\,d\beta, \\ &= -\int_1^{-1} e^{-q^2 D_{\|}tx^2} J_p^2\left(aq\sqrt{1-x^2}\right)dx, \\ &= \int_{-1}^{1} e^{-q^2 D_{\|}tx^2} J_p^2\left(aq\sqrt{1-x^2}\right)dx, \\ &= 2\int_0^{1} e^{-q^2 D_{\|}tx^2} J_p^2\left(aq\sqrt{1-x^2}\right)dx. \end{aligned} \tag{47}$$

We introduced the change of variables $x=\cos(\beta)$, thus, $\sin(\beta)=\sqrt{1-x^2}$ and $\sin(\beta)\,d\beta = -d\cos(\beta) = -dx$. The integral is symmetric around zero; therefore, we integrate from zero to one.

The above integral does not have a compact closed-form solution. However, it can be solved by expanding the squared Bessel function of the first kind in series [119]:

$$J_p^2(z) = \sum_{k=0}^{\infty} c_{kp} z^{2(p+k)}, \tag{48}$$

where the coefficients $c_{kp}$ are determined by

$$c_{kp} = \frac{(-1)^k}{k!(2p+k)!}\left(\frac{1}{2}\right)^{2(p+k)}\binom{2(p+k)}{p+k}. \tag{49}$$

Inserting Eq. (48) into Eq. (47) we obtain

$$I(p)=2\sum_{k=0}^{\infty}c_{kp}\left(aq\right)^{2(p+k)}\int_0^1 e^{-q^2D_{\|}tx^2}\left(1-x^2\right)^{(p+k)}dx. \tag{50}$$

Using the binomial theorem,

$$\left(1-x^2\right)^{(p+k)}=\sum_{j=0}^{p+k}\binom{p+k}{j}(-1)^j x^{2j}\ , \tag{51}$$

we obtain,

$$I(p)=2\sum_{k=0}^{\infty}c_{kp}\left(aq\right)^{2(p+k)}\sum_{j=0}^{p+k}\binom{p+k}{j}(-1)^j\int_0^1 e^{-q^2D_{\|}tx^2}x^{2j}dx\,. \tag{52}$$

The integral in the previous equation can be solved as

$$\int_0^1 e^{-q^2D_{\|}tx^2}x^{2j}dx=\frac{1}{2\left(q^2D_{\|}t\right)^{j+1/2}}\left(\Gamma\left(j+\frac{1}{2}\right)-\Gamma\left(j+\frac{1}{2},q^2D_{\|}t\right)\right), \tag{53}$$

where $\Gamma\left(j+1/2,bD\right)$ is the upper incomplete Gamma function. It is important to note that this function has been implemented in various libraries using different formats. For example, in *scipy*, it is defined by $\Gamma\left(j+1/2,x\right)=\Gamma\left(j+1/2\right)gammaincc\left(j+1/2,x\right)$, and by $igamma$ in *Matlab*.

By plugging Eq. (53) into Eq. (52) we get

$$I(p)=\sum_{k=0}^{\infty}c_{kp}\left(aq\right)^{2(p+k)}\sum_{j=0}^{p+k}\binom{p+k}{j}(-1)^j\frac{\Gamma\left(j+\frac{1}{2}\right)-\Gamma\left(j+\frac{1}{2},q^2D_{\|}t\right)}{\left(q^2D_{\|}t\right)^{j+1/2}}. \tag{54}$$

Inserting this result into Eq. (46),

$$\begin{aligned}\left\langle S\right\rangle=\frac{E\left(\mathbf{q}=0,t\right)}{2}&\left[\sum_{k=0}^{\infty}c_{k0}\left(aq\right)^{2k}\sum_{j=0}^{k}\binom{k}{j}(-1)^j\frac{\Gamma\left(j+\frac{1}{2}\right)-\Gamma\left(j+\frac{1}{2},q^2D_{\|}t\right)}{\left(q^2D_{\|}t\right)^{j+1/2}}\right.\\&\left.+\sum_{p=1}^{\infty}e^{-p^2\frac{D_{\|}t}{a^2}}\sum_{k=0}^{\infty}c_{kp}\left(aq\right)^{2(p+k)}\sum_{j=0}^{p+k}\binom{p+k}{j}(-1)^j\frac{\Gamma\left(j+\frac{1}{2}\right)-\Gamma\left(j+\frac{1}{2},q^2D_{\|}t\right)}{\left(q^2D_{\|}t\right)^{j+1/2}}\right],\end{aligned} \tag{55}$$

Substituting Eq. (49) into Eq. (55) and considering Lori's q-space correction, we obtain the final spherical mean signal model in Eq. (17).

Note that for $q=0$ the previous equation exhibits a singularity due to the division by $q$. To address this issue, the following asymptotic limit for $q\to 0$ is employed to avoid numerical issues: $\frac{\Gamma\left(j+\frac{1}{2}\right)-\Gamma\left(j+\frac{1}{2},q^2D_{\|}t\right)}{\left(q^2D_{\|}t\right)^{j+1/2}}=\frac{\Gamma\left(j+\frac{1}{2}\right)}{\Gamma\left(j+\frac{3}{2}\right)}=\frac{1}{j+\frac{1}{2}}$.

**Appendix E: Derivation of the distribution of cylinder radius**

To solve Eq. (25), let us focus on the integral

$$I=\int_{a\cdot g}^{a/g} a_i^{\mu-2}e^{-\kappa a_i}da_i=\frac{1}{\kappa^{\mu-1}}\int_{\kappa\cdot a\cdot g}^{\kappa\cdot a/g} v^{\mu-2}e^{-v}dv\,, \tag{56}$$

where we used the substitution $v=\kappa a_i$, thus, $da_i=dv/\kappa$.

We recognize this integral can be written in terms of the upper incomplete Gamma function,

$$\Gamma\left(s,x\right)=\int_{x}^{\infty}t^{s-1}e^{-t}dt\,. \tag{57}$$

However, since our limits of integration are from $\kappa\cdot a\cdot g$ to $\kappa\cdot a/g$, we need to use the general form:

$$\int_{a}^{b}t^{s-1}e^{-t}dt=\Gamma\left(s,a\right)-\Gamma\left(s,b\right). \tag{58}$$

In our case, we get

$$I=\frac{1}{\kappa^{\mu-1}}\left[\Gamma\left(\mu-1,a\cdot g\cdot\kappa\right)-\Gamma\left(\mu-1,\frac{a\cdot\kappa}{g}\right)\right]. \tag{59}$$

On the other hand, the normalization constant $\eta$ is estimated from $\int_0^\infty P\left(a\right)da=1$, thus

$$\eta\frac{\kappa}{\Gamma\left(\mu\right)}\frac{g}{\left(1-g\right)}\left[\int_0^\infty\Gamma\left(\mu-1,a\cdot g\cdot\kappa\right)da-\int_0^\infty\Gamma\left(\mu-1,\frac{a\cdot\kappa}{g}\right)da\right]=1, \tag{60}$$

where we inserted the result in Eq. (59) in Eq. (25).

Making the substitution $x = a \cdot g \cdot \kappa$ and thus $dx = g \cdot \kappa da$ in the first integral and $x = a \cdot \kappa / g$, $dx = (\kappa / g) da$ in the second one, we obtain

$$\eta \frac{\kappa}{\Gamma(\mu)} \frac{g}{(1-g)} \left( \frac{1}{g \cdot \kappa} - \frac{g}{\kappa} \right) \int_0^\infty \Gamma(\mu - 1, x) dx = 1. \tag{61}$$

The integral in the last equation is equal to $\Gamma(\mu)$ [119], and thus,

$$\eta = \frac{(1-g)}{(1-g^2)}. \tag{62}$$

Substituting Eqs. (62) and (59) into Eq. (25), we obtain the final solution.